\documentclass[aps,pre,showpacs,showkeys,onecolumn,superscriptaddress]{revtex4}
\usepackage[dvips]{graphicx}
\usepackage[dvips]{color}
\usepackage{hyperref,breakurl,amsmath,amssymb,pst-node,eepic}

\newcommand{\be}{\begin{equation}}
\newcommand{\ee}{\end{equation}}

\begin{document}

\title{Commutative law for products of infinitely large isotropic random matrices}

\author{Zdzislaw Burda}
\email{zdzislaw.burda@uj.edu.pl}
\affiliation{Marian Smoluchowski Institute of Physics,
Jagiellonian University, Reymonta 4, 30--059 Krak\'{o}w, Poland}
\affiliation{Mark Kac Complex Systems Research Centre,
Jagiellonian University, Reymonta 4, 30--059 Krak\'{o}w, Poland}

\author{Giacomo Livan}
\email{glivan@ictp.it}
\affiliation{Abdus Salam International Centre for Theoretical Physics, Strada Costiera 11, 34151 Trieste, Italy}

\author{Artur Swiech}
\email{artur.swiech@uj.edu.pl}
\affiliation{Marian Smoluchowski Institute of Physics, 
Jagiellonian University, Reymonta 4, 30--059 Krak\'{o}w, Poland}


\begin{abstract}
Ensembles of isotropic random matrices are defined by the invariance of the probability measure under the left (and right) multiplication by an arbitrary unitary matrix.
We show that the multiplication of large isotropic random matrices is spectrally commutative and self-averaging in the limit of infinite matrix size $N\rightarrow \infty$. The notion of spectral commutativity means that the eigenvalue density of a product $ABC\ldots$ of such matrices is independent of the order of matrix multiplication, for example the matrix $ABCD$ has the same eigenvalue density as $ADCB$. In turn, the notion of self-averaging means that the product of $n$ independent but identically distributed random matrices, which we symbolically denote by $AAA\ldots$, has the same eigenvalue density as the corresponding power $A^n$ of a single matrix drawn from the underlying matrix ensemble. For example, the eigenvalue density of $ABCCABC$ is the same as of $A^2B^2C^3$. We also discuss the singular behavior of the eigenvalue and singular value densities of isotropic matrices and their products for small eigenvalues $\lambda \rightarrow 0$. We show that the singularities at the origin of the eigenvalue density and of the singular value density are in one-to-one correspondence in the limit $N\rightarrow \infty$: the eigenvalue density of an isotropic random matrix has a power law singularity at the origin $\sim |\lambda|^{-s}$ with a power $s\in(0,2)$ when and only when the density of its singular values has a power law singularity $\sim \lambda^{-\sigma}$ with a power $\sigma=s/(4-s)$. These results are obtained analytically in the limit $N\rightarrow \infty$. We supplement these results with numerical simulations for large but finite $N$ and discuss finite size effects for the most common ensembles of isotropic random matrices. 



\end{abstract}

\maketitle

\section{Introduction}
\label{sec:intro}

Ensembles of Hermitian random matrices with invariant measures have been thoroughly studied in the literature \cite{m,gmw,agz,abd}. Much less known are non-Hermitian random matrices \cite{ks}. In this paper we discuss a class of isotropic non-Hermitian matrices that represent a natural extension of the class of invariant Hermitian matrices to the non-Hermitian case: the probability measure of an isotropic random matrix ensemble is invariant under the left (and right) multiplication by an arbitrary unitary matrix. Here we are interested in properties of the limiting eigenvalue densities of products of isotropic matrices in the limit of infinite matrix size $N\rightarrow \infty$. These properties can be deduced from the correspondence between large random matrices and free random variables \cite{vdn}, and most of them follow from the Haagerup-Larsen theorem \cite{hl}, that was formulated in the framework of free probability. This theorem gives a very useful relation between the eigenvalue density of an isotropic matrix $A$ and the density of its invariant Hermitian partner $AA^\dagger$. We exploit this relation to discuss the spectral commutativity and the self-averaging of the product of isotropic random matrices in the large $N$ limit. The product of identically distributed independent matrices has the same eigenvalue density as the corresponding power of a single random matrix \cite{bns}. This is an exceptional property which has no counterpart in classical probability theory.

The paper is organized as follows. In Section \ref{sec:irm} we recall the definition of isotropic random matrices and the Haagerup-Larsen theorem \cite{hl}. In Section \ref{sec:prod} we discuss products of isotropic matrices and the spectral commutativity of the multiplication of infinitely large matrices from this class. In section \ref{sec:examples} we illustrate how to use the Haagerup-Larsen relation to calculate the eigenvalue density for a few isotropic matrices in the large $N$-limit. In Section \ref{sec:sv} we analyze the correspondence between the singularities of the eigenvalue density and the singular value density. In Section \ref{sec:fs} we discuss finite size effects for three generic ensembles of isotropic matrices. In Section \ref{sec:fsp} we compare eigenvalue densities for products of finite matrices, obtained by Monte-Carlo simulations, with the limiting densities calculated analytically using the method discussed earlier in Section \ref{sec:prod}. In Section \ref{sec:discussion} we shortly summarize the paper. 

\section{Isotropic random matrices}
\label{sec:irm}

Before we discuss isotropic matrices, let us recall invariant Hermitian random matrices. A random Hermitian matrix $h$ is called invariant if its probability measure is invariant under the transformation $h \rightarrow U^{-1} h U$ where $U$ is an arbitrary unitary matrix. The notion of random matrix is analogous to random variable, so one has to remember that the term ``random matrix'' does not refer to a single instance but to an ensemble of matrices with a given probability measure. Thus, invariance of $h$ stands for the invariance of the probability measure or, in other words, that the random matrices $h$ and $U^{-1} h U$ have the same probability measures. Randomness of an invariant Hermitian matrix is entirely encoded in its eigenvalue distribution in contrast to non-invariant random matrices. However, with any Hermitian non-invariant random matrix $h_{n}$ one can associate a unique invariant random matrix $h$ that has exactly the same eigenvalue distribution as $h_{n}$: $h=uh_{n} u^{-1}$ where $u$ is drawn according to the uniform (Haar) measure from the unitary group $U(N)$.

A random matrix $H$ is called isotropic \cite{bns} if it can be decomposed into a product of an invariant Hermitian positive semi-definite random matrix $h$ and a Haar unitary matrix $u$: $H=h u$. Clearly $h^2 = HH^\dagger$. Statistical properties of $H$ are inherited from its invariant Hermitian partner $h$. Eigenvalues of $h$ correspond to singular values of $H$. The concept of isotropic random matrices is a natural generalization of the concept of rotationally invariant complex variables that can be written as $z=r e^{i\phi}$, where $r$ is a non-negative real random variable and $\phi$ is a random variable uniformly distributed on the interval $[0,2\pi)$.

One should note that isotropic matrices can be constructed also from non-invariant random Hermitian matrices as $H=u_1 h_{n} u_2$, where $u_1,u_2$ are independent unitary Haar matrices and $h_{n}$ is an arbitrary Hermitian positive semi-definite random matrix which is not necessarily invariant. In particular $h_{n}$ may be a diagonal random matrix with independent identically distributed non-negative real random variables on the diagonal.

The probability measure of an isotropic random matrix is invariant under the right $H\rightarrow HU$ (and left $H\rightarrow UH$) multiplication by any unitary matrix $U$. This property can be used as an alternative definition of isotropic random matrices. The eigenvalue distribution $\rho_H(z)$ of an isotropic random matrix $H$ is circularly symmetric on the complex plane. It depends only on the eigenvalue modulus $r=|z|$, so it can be written as a function of a single real argument $\rho_H(z) = \varrho_H(|z|)$. The radial profile of the eigenvalue distribution of $H$ depends on the eigenvalue distribution $\rho_h(\lambda)$ of the matrix $h$ and on the size $N$ of the matrix. In two limiting cases of matrix dimensions, $N=1$ and $N\rightarrow \infty$, the relation between eigenvalue distribution of the isotropic matrix $H$ and of its Hermitian partner $h$ is explicitly known. 

For $N=1$ the random matrix $H$ reduces to a complex random variable, and the matrix $h$ to a real non-negative random variable. The relation between the two reads $H=h e^{i\phi}$ where $\phi$ is a phase uniformly distributed on the interval $[0,2\pi)$. The relation between the distributions of the random variables $H$ and $h$ can be conveniently expressed in terms of the cumulative density function $F_h(x)=\int_0^x \rho_h(\lambda) d\lambda $ for $h$ and the radial cumulative distribution function for $H$
\begin{equation}
\label{eq:F}
F_H(x) = \int_{|z|\le x} \rho_H(z) d^2 z = \int_{|z|\le x} \varrho_H(|z|) d^2 z =
\int_0^x 2\pi r \varrho_H(r) dr \ . 
\end{equation}
The relation between the cumulative distributions for $H$ and $h$ reads $F_H(x)=F_h(x)$. It amounts to $2\pi r \varrho_H(r) = \rho_h(r)$. Of course this relation holds only for $N=1$.

Below we use the radial cumulative density function $F_H(x)$ also for $N>1$. In this case $F_H(x)$ is defined as the probability that a randomly chosen eigenvalue of $H$ lies within distance $x$ from the origin of the complex plane.  

As we mentioned, the relation between the eigenvalue distributions of $H$ and $h$ is also known for $N\rightarrow \infty$ \cite{hl}, as in this limit random matrices can be mapped onto free random variables \cite{ns1}: invariant Hermitian matrices to free real random variables, and isotropic random matrices to R-diagonal free random variables \cite{ns2}. In other words, in this case one can use methods of free probability \cite{vdn} to derive the relation. We quote here the result that is known as the Haagerup-Larsen theorem \cite{hl}. This theorem states that the radial cumulative density function of $H$ can be expressed in terms of the S-transform \cite{v,vdn} for $h^2 = HH^\dagger$
\begin{equation}
\label{eq:HLtheorem}
S_{h^2}\left(F_H(x)-1\right)=\frac{1}{x^{2}} \ . 
\end{equation}
The support of the radial profile $\varrho_H(r)$ extends from $r_{min}$ to $r_{max}$, so that the eigenvalue density of $H$ forms either a disk of radius $r_{max}$ if $r_{min}=0$ or a ring if $r_{min}>0$. The disc (or ring) is centered at the origin of the complex plane. The internal radius is $r^{-2}_{min}= \int \rho_h(\lambda) \lambda^{-2} d\lambda$ and the external one $r^2_{max}= \int \rho_h(\lambda) \lambda^2 d\lambda$. The internal radius is positive $r_{min}>0$ and the support of $\rho_H$ is a ring, if for $\lambda\rightarrow 0$ the eigenvalue density $\rho_h(\lambda)$ decays to zero faster than the first power of $\lambda$: $\rho_h(\lambda)\sim \lambda^{1+\epsilon}$, $\epsilon>0$. Otherwise, the support of the eigenvalue density of $H$ is a disk. To be precise, the theorem assumes that the eigenvalue density of $h$ has no isolated point masses (Dirac deltas) in the spectrum. In other words, the spectrum of $h$ must be continuous.

Eq. (\ref{eq:HLtheorem}) tells us that the cumulative distribution $F_H$ implicitly depends on the eigenvalue density of the matrix $h$ through the S-transform $S_{h^2}$. So, let us recall what the S-transform is \cite{v,vdn}. It is defined for an infinitely large ($N\rightarrow \infty)$ invariant Hermitian matrix. Let $a$ be such a matrix. Denoting the limiting eigenvalue density of this matrix by $\rho_a(\lambda)$, the S-transform is calculated as follows. First, one calculates the Green's function as the Stieltjes transform of the eigenvalue density 
\begin{equation}
\label{eq:G}
G_a(z)=\int_I \frac{\rho_a(\lambda) d\lambda}{z-\lambda} .
\end{equation}
$G_a$ is a complex function defined outside the support $I$ of the eigenvalue density, which consists of intervals on the real axis. The Green's function can be expanded in powers of $1/z$, and the coefficients of this expansion are equal to the moments of the eigenvalue distribution:
\begin{equation}
\mu_{a,k} = \int_I \rho_a(\lambda) \lambda^k d\lambda \ ,
\end{equation}
for $k=0,1\ldots$ and $\mu_{a,0}=1$. One can alternatively define the 
moment-generating function: 
\begin{equation}
\label{eq:phi}
\phi_a(z) = \frac{1}{z} G_a\left(\frac{1}{z}\right) - 1 \ .
\end{equation}  
Expanding it in $z$ one obtains an infinite power series 
$\phi_a(z) =\sum_{k=1}^\infty \mu_{a,k} z^k$ if all moments exist. 
The S-transform for the matrix $a$ is defined as 
\begin{equation}
\label{eq:S_chi}
S_a(z) = \frac{z+1}{z} \chi_a(z) \ ,
\end{equation} 
where $\chi_a$ is the inverse of the moment-generating function $\phi_a$:
\begin{equation}
\label{eq:inverse}
\chi_a(\phi_a(z)) = \phi_a(\chi_a(z)) = z \ .
\end{equation}
The S-transform of the product of independent invariant matrices $a,b$ (free random variables) is equal to the product of the corresponding S-transforms \cite{v}:
\begin{equation}
\label{eq:Sab}
S_{ab}(z) = S_{a}(z) S_b(z) \ .
\end{equation}
Since the S-transform is a complex-valued function the multiplication on the right hand side of the equation is both associative and commutative. This property has deep consequences for products of isotropic matrices in the limit $N\rightarrow \infty$. In particular, as we discuss in the next section, multiplication of isotropic random matrices is commutative in this limit.

Coming back to Eq. (\ref{eq:HLtheorem}) we see that in order to calculate the cumulative distribution of an isotropic matrix $H$, we first have to calculate the S-transform for the matrix $h^2 =HH^\dagger$. Assume we know the eigenvalue density of $h$. The eigenvalue density of $h^2$ is
\begin{equation}
\label{eq:rho_h2}
\rho_{h^2}(\lambda) = \frac{1}{2\sqrt{\lambda}} \rho_h\left(\sqrt{\lambda}\right) \ .
\end{equation}
Finding the Green's function $G_{h^2}(z)$ for $h^2$ as in Eq. (\ref{eq:G}) and then the S-transform $S_{h^2}(z)$ as in Eq. (\ref{eq:S_chi}), we obtain an explicit equation for $F_H$ (Eg. \ref{eq:HLtheorem}). Actually, if one eliminates the S-transform from Eq. (\ref{eq:HLtheorem}) one can rewrite the latter as 
\begin{equation}
F_H(x) - 1 = x^2 G_{h^2}\left(\frac{x^2 F_H(x)}{F_H(x)-1}\right) \ ,
\end{equation}
applying to it Eqs. (\ref{eq:phi},\ref{eq:S_chi}). This equation has been derived independently in Refs. \cite{fz,fsz}. This form is however less transparent than Eq. (\ref{eq:HLtheorem}), which explicitly refers to the S-transform and thus uncovers an important connection to the commutative nature of the multiplication of the S-transform (Eq. \ref{eq:Sab}) that is responsible for the spectral commutativity of the isotropic matrices in the large $N$-limit, as we discuss in the next section.

\section{Products of isotropic random matrices}
\label{sec:prod}

Consider a product of a finite number, $n$, of independent isotropic random matrices
\begin{equation}
\label{eq:prod}
P = H_1 H_2 \ldots H_n
\end{equation}
in the limit $N\rightarrow \infty$. The goal is to calculate the eigenvalue density of $P$ given the eigenvalue densities of the $H_i$'s. The product of isotropic random matrices is also isotropic so the eigenvalue density of the product can be determined from Eq. (\ref{eq:HLtheorem}) replacing $H$ by $P$ and $h$ by $p$. So we have to calculate
the S-transform for $p^2=PP^\dagger$. As a consequence of the multiplication law (\ref{eq:Sab}) the S-transform of the matrix for $p^2=PP^\dagger$ with $P$ given by Eq. (\ref{eq:prod}) can be written as a product of S-transforms \cite{bjlns1,bjlns2}:
\begin{equation}
S_{p^2}(z) = \prod_{j=1}^n S_{h_j^2}(z) 
\end{equation}
for $h_j^2=H_j H_j^\dagger$. It is a simple consequence of the associativity and commutativity of the S-transform composition rule (\ref{eq:Sab}). This means that the equation for the radial cumulative eigenvalue density $F_P(x)$ of the matrix $P$ (\ref{eq:prod}) can be expressed in terms of the S-transforms for individual factors in the product (\ref{eq:prod}):
\begin{equation}
\label{eq:S1n}
\prod_{j=1}^n S_{h_j^2}\left(F_P(x)-1\right) = \frac{1}{x^2} \ .
\end{equation}
If one writes an analogous equation for the product
\begin{equation}
\label{eq:prod_pi}
P_\pi = H_{\pi(1)} H_{\pi(2)} \ldots H_{\pi(n)} ,
\end{equation}
where $\pi$ is an arbitrary permutation of the set $\{1,\ldots,n\}$ one obtains
exactly the same equation as in (\ref{eq:S1n})
\begin{equation}
\prod_{j=1}^n S_{h_j^2}\left(F_{P_\pi}(x)-1\right) = \frac{1}{x^2} \ ,
\end{equation}
but now for $F_{P_\pi}$. Therefore the functions $F_P$ and $F_{P_\pi}$ are identical $F_P\equiv F_{P_\pi}$. In other words, the eigenvalue distribution of the product of isotropic matrices (\ref{eq:prod_pi}) does not depend on the order of matrix multiplication in the large $N$-limit. 

As we mentioned, the multiplication of infinitely large isotropic matrices has another surprising property. The product of $n$ independent identically distributed matrices isotropic that we denote by $P=H H \ldots H$ has exactly the same probability measure as a power $Q=H^n$ of a single matrix \cite{bns}. It was first discovered for the product of Ginibre matrices \cite{bjw}. For the sake of completness we repeat here the argument given in Ref. \cite{bns}. Eq. (\ref{eq:S1n}) for the product of identically distributed matrices takes the form
\begin{equation}
S_{h^2}(F_P(x)-1) = \frac{1}{x^{2/n}} \ ,
\end{equation}
which means that $F_P(x) = F_H\left(x^{1/n}\right)$. On the other hand, by construction, the eigenvalues of $Q$ are given by powers of eigenvalues of $H$: $\lambda_Q=\lambda_H^n$, so the probability that $|\lambda_Q|<x$ is equal to the probability that $|\lambda_H|^n<x$: $\mbox{Prob}(|\lambda_Q|<x) = \mbox{Prob}(|\lambda_H|^n<x)$. It follows that $F_Q(x)=F_H\left(x^{1/n}\right)$ and hence $F_P(x)=F_Q(x)$.

To summarize this section, the product of a finite number of independent infinitely dimensional isotropic random matrices is an isotropic random matrix. The eigenvalue distribution of this matrix does not depend on the order of multiplication. Moreover, if there are independent identically distributed matrices in the product they can be substituted by the corresponding power of a single random matrix from the corresponding matrix ensemble. For instance the product $ACBACCB$ has the same limiting eigenvalue density as $A^2 B^2 C^3$. It is worth mentioning that the effect of self-averaging was also observed for a Wigner type of matrices with independent identically distributed. entries belonging to the Gaussian universality class \cite{agt}. Intuitively, such Wigner matrices become isotropic in the large $N$ limit.

\section{Examples of infinitely dimensional isotropic random matrices}
\label{sec:examples}

In this section we give a couple of examples of isotropic matrices in the limit $N\rightarrow \infty$. We keep the convention that isotropic matrices are denoted by capital letters and their Hermitian partners by the corresponding small letters.

\begin{figure}
\centering
\includegraphics[width=0.4\textwidth]{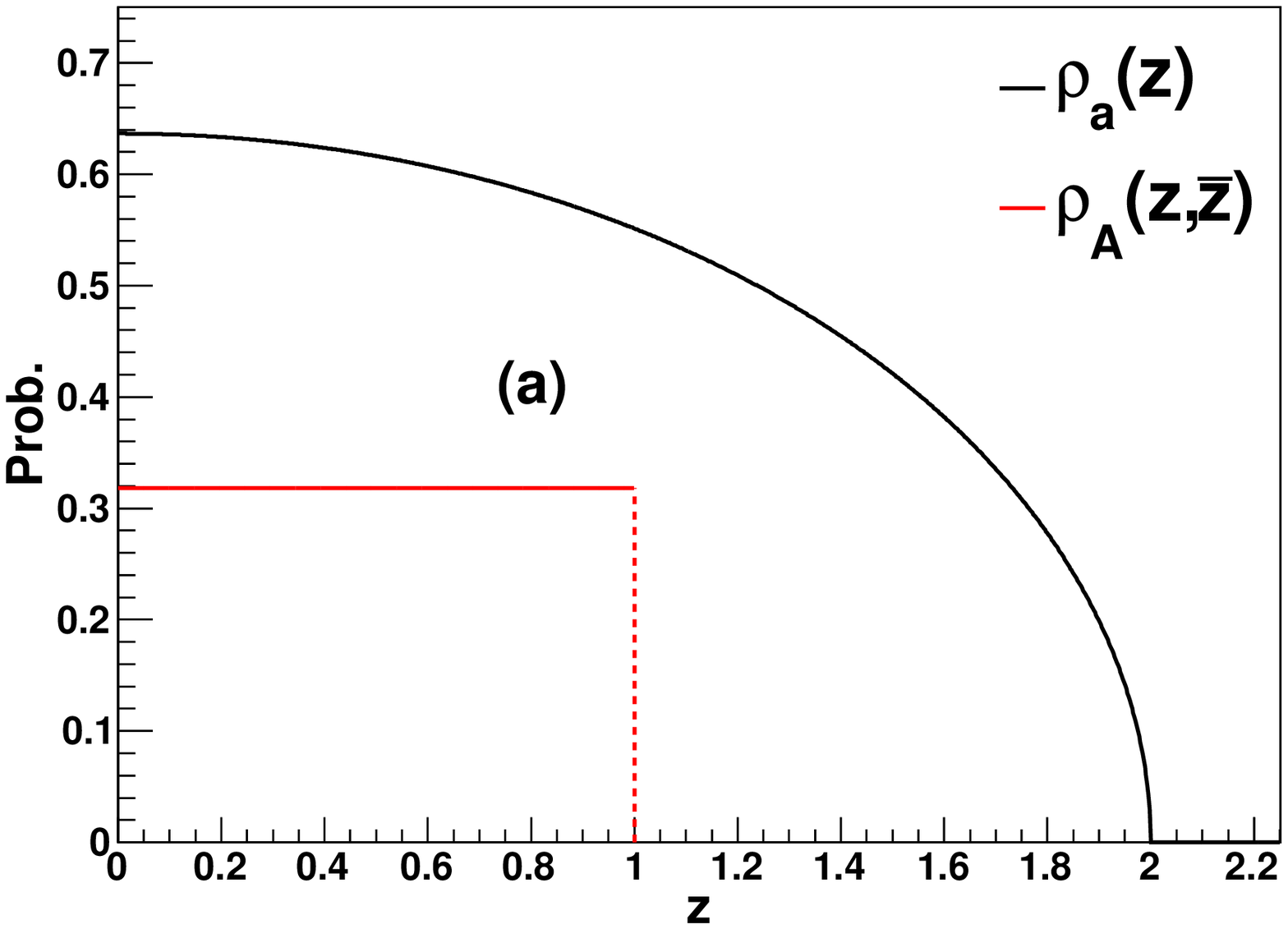}
\includegraphics[width=0.4\textwidth]{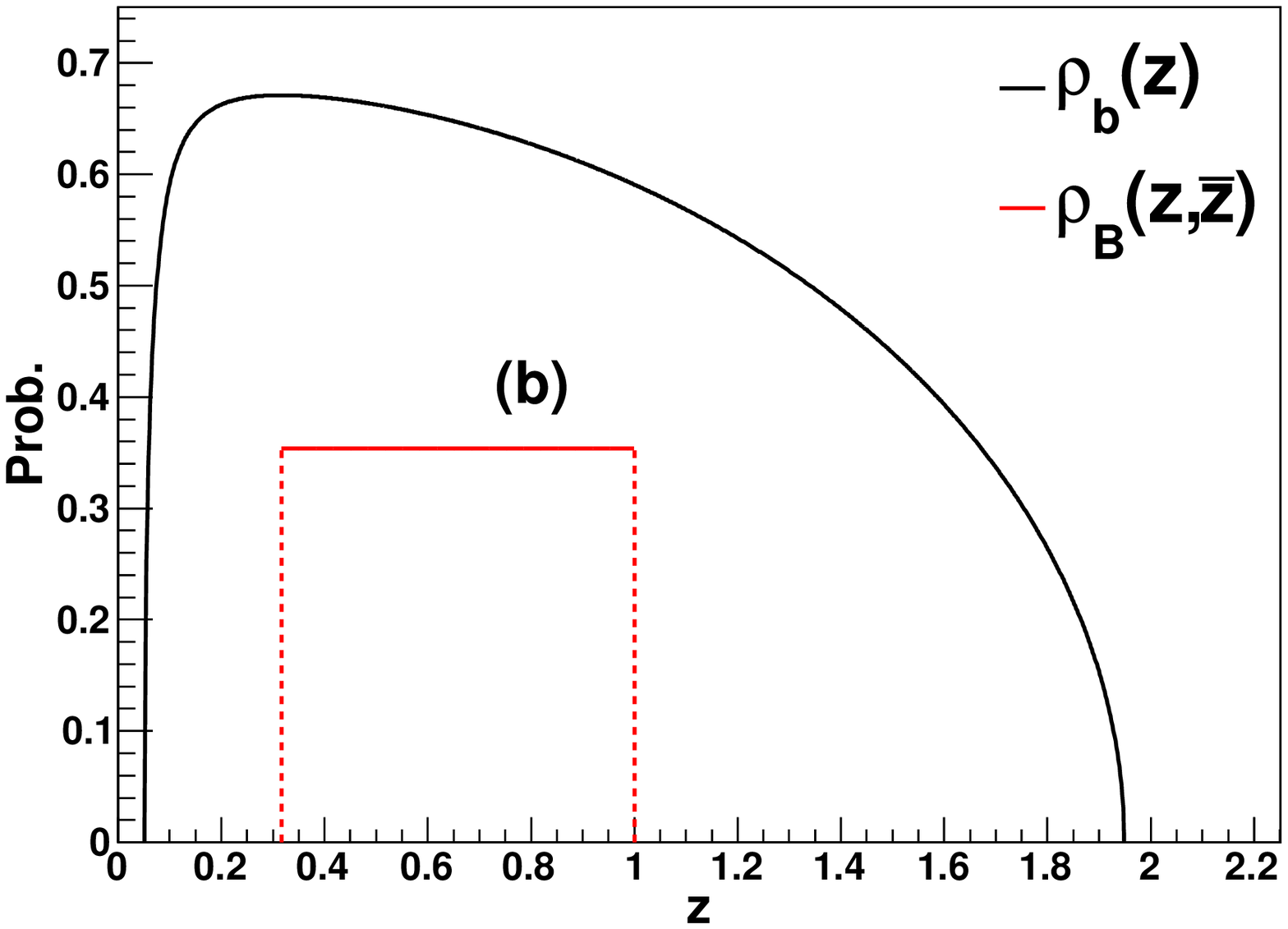}\\
\includegraphics[width=0.4\textwidth]{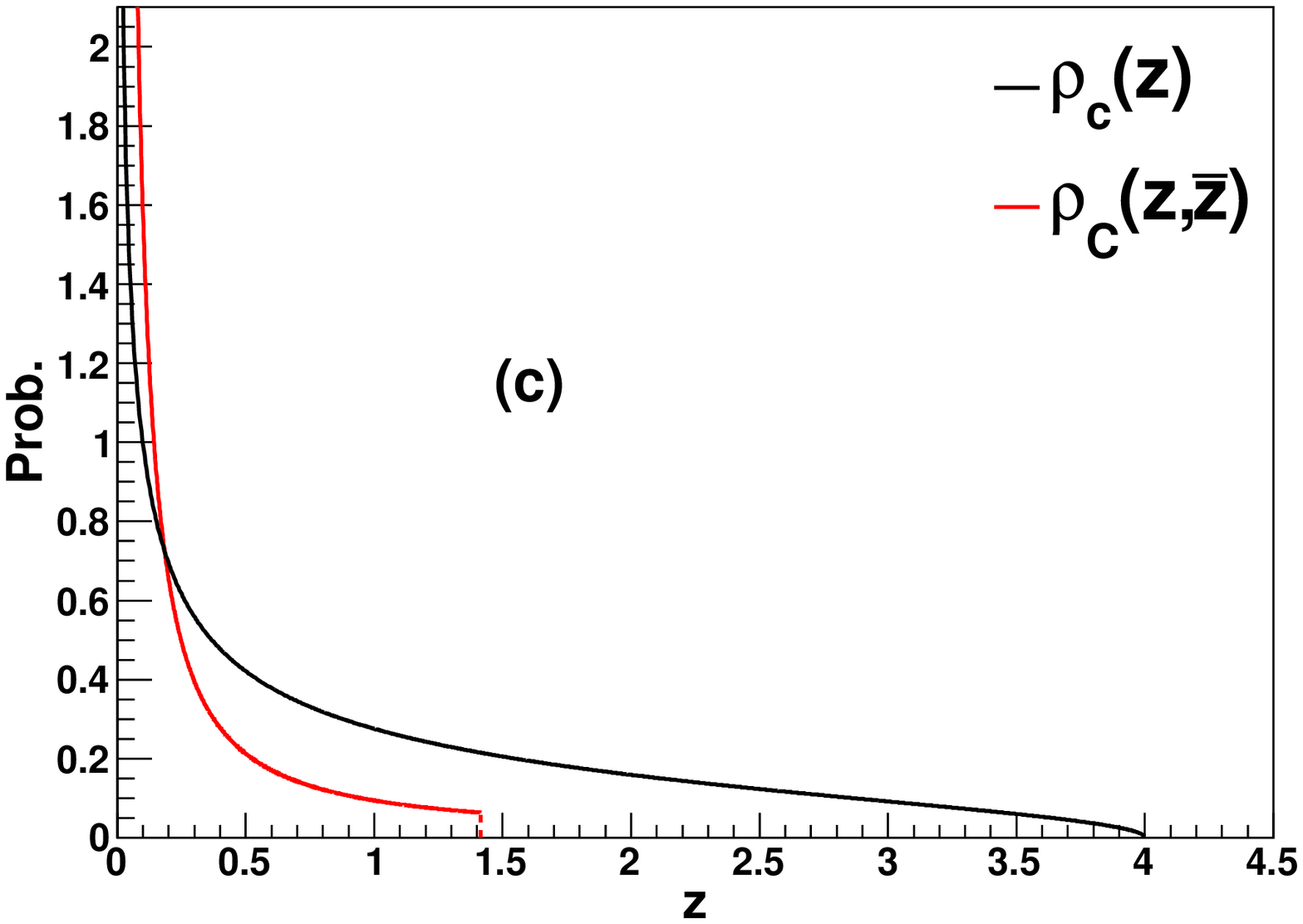}
\includegraphics[width=0.4\textwidth]{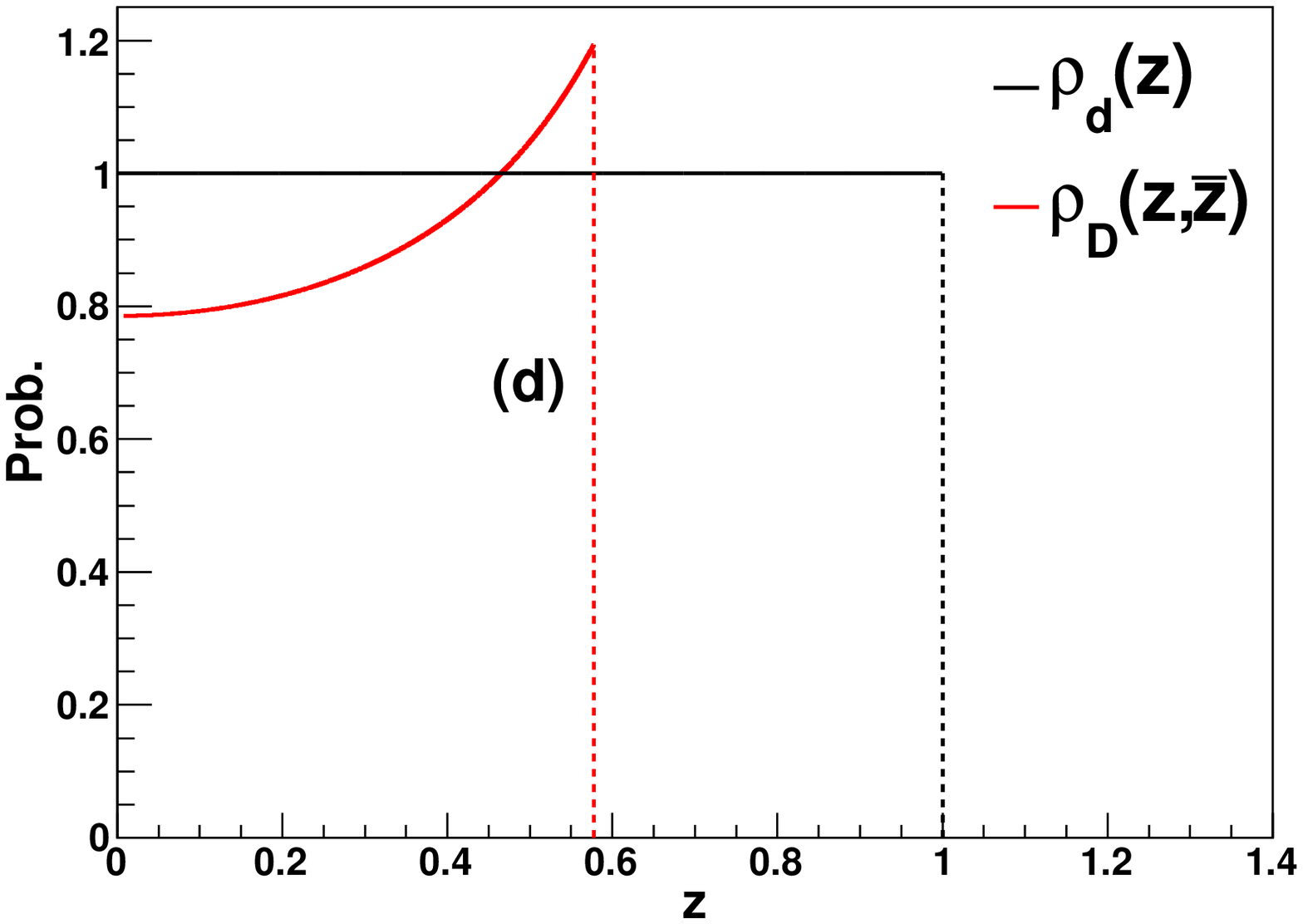}
\caption{Radial profile (red) of the eigenvalue density $\rho(z)$ for isotropic random matrices matrices of the type $A$, $B$, $C$ and $D$ discussed in Section \ref{sec:examples} and the corresponding eigenvalue densities (black) for their Hermitian partners $a$, $b$, $c$ and $d$.}
\label{fig:SingleMatrices}
\end{figure}

The first example is a matrix $A=au$, where $u$ is a Haar unitary random matrix
and $a$ is an invariant Hermitian random matrix with an eigenvalue density given by the quarter-circle law:
\begin{equation}
\label{eq:QC}
\rho_a(\lambda) = \frac{1}{\pi} \sqrt{4-\lambda^2} \ , \ \lambda \in [0,2] \ .
\end{equation}
The matrix $a^2$ has the eigenvalue density (\ref{eq:rho_h2}):
\begin{equation}
\label{eq:W}
\rho_{a^2}(\lambda) = \frac{1}{2\pi} \sqrt{\frac{4-\lambda}{\lambda}} \ , \
\lambda \in [0,4] \ .
\end{equation}
The Green's function (\ref{eq:G}) of $a^2$ is
\begin{equation}
G_{a^2}(z) = \frac{1}{2} - \frac{1}{2} \sqrt{\frac{z-4}{z}} \ .
\end{equation}
Using Eq. (\ref{eq:phi}) we find the moment-generating function
\begin{equation}
\phi_{a^2}(z) = \frac{1}{2z}\left(1 - \sqrt{1-4z}\right) - 1 
\end{equation}
and its inverse function (\ref{eq:inverse}) 
\begin{equation}
\chi_{a^2}(z) = \frac{z}{(z+1)^2} \ .
\end{equation}
The S-transform (\ref{eq:S_chi}) reads
\begin{equation}
\label{eq:Sa2}
S_{a^2}(z) = \frac{1}{1+z} \ .
\end{equation}
Inserting it into Eq. (\ref{eq:HLtheorem}) we find
the radial cumulative distribution for the isotropic matrix $A$ associated with $a$: 
\begin{equation}
F_{A}(x) = x^2
\end{equation}
for $0\le x\le 1$.  The radial profile is $\varrho_A(r) = \frac{1}{2\pi r} F'_A(r) = \frac{1}{\pi}$, and hence
\begin{equation}
\label{eq:Ginibre}
\rho_A(z) = \frac{1}{\pi} \ , \ |z|\le 1 .
\end{equation}
The eigenvalue density $\rho_a(x)$ of the Hermitian matrix $a$ and the corresponding radial profile $\varrho_A(x)$ of the eigenvalue density of the isotropic matrix $A=au$ are shown in Fig. \ref{fig:SingleMatrices}a. Clearly in this case the spectrum of the matrix $A$ is the same as for Ginibre matrices \cite{g1}.

As a second example, we consider a random isotropic matrix $B=bu$ associated with an invariant Hermitian random matrix $b$ with the following eigenvalue density
\begin{equation}
\rho_b(\lambda) = \frac{1}{\pi \alpha \lambda} 
\sqrt{\left(\alpha^2_+ - \lambda^2)(\lambda^2-\alpha^2_-\right)} \ , \
\lambda \in [\alpha_-,\alpha_+] \ ,
\end{equation}
where $\alpha_\pm = 1 \pm \sqrt{\alpha}$  and $0<\alpha\le 1$.  
For $b^2$ we have
\begin{equation}
\rho_{b^2}(\lambda) = \frac{1}{2\pi \alpha \lambda} 
\sqrt{\left(\lambda_+ - \lambda)(\lambda-\lambda_-\right)} \ , \
\lambda \in [\lambda_-,\lambda_+] 
\end{equation}
where $\lambda_\pm = (1\pm \sqrt{\alpha})^2$.
It is straightforward to calculate the Green's function:
\begin{equation}
G_{b^2}(z) = \frac{1}{2\alpha z} \left(z + \alpha -1 - \sqrt{(z-\lambda_+)(z-\lambda_-)}\right) \ ,
\end{equation}
the moment generating function
\begin{equation}
\phi_{b^2}(z) = \frac{1}{2\alpha z}\left(1 - \sqrt{(1-\lambda_+z)(1-\lambda_-z)}\right) - \frac{\alpha+1}{2\alpha} \ ,
\end{equation}
its inverse function
\begin{equation}
\chi_{b^2}(z) = \frac{z}{(z+1)(\alpha z + 1)} \ ,
\end{equation}
and the $S$-transform for the matrix $b^2$ 
\begin{equation}
\label{eq:Sb2}
S_{b^2}(z) = \frac{1}{\alpha z + 1} \ .
\end{equation}
Using the Haagerup-Larsen theorem (\ref{eq:HLtheorem}) 
we find a very simple equation for the radial cumulative distribution $F_B(r)$
of the matrix $B$
\begin{equation}
F_B(r) = \frac{r^2 - r_0^2}{1-r_0^2}  \ , \ r \in [r_0,1] \ ,
\end{equation}
where $r_0 = \sqrt{1-\alpha}$. It corresponds to a uniform eigenvalue distribution in the ring with the internal radius $r_{min}=r_0$ and the external one $r_{max}=1$ with the density
\begin{equation}
\rho_B(z) = \frac{1}{\pi(1-r_0^2)} \ , \ |z| \in [r_0,1] \ . 
\end{equation} 
Clearly for $\alpha=1$ the previous case is restored. In the remaining part of the paper we choose $\alpha=0.9$ for the matrix $B$ unless stated otherwise. For this choice $r_{min}=\sqrt{0.1}$ and $\rho_B(z) = \frac{1}{0.9\pi}$ for $|z| \in [\sqrt{0.1},1]$ (see Fig. (\ref{fig:SingleMatrices}b)) 

Next, we define a matrix $C=cu$ with $c$ having the following eigenvalue density
\begin{equation}
\rho_c(\lambda) = \frac{1}{2\pi} \sqrt{\frac{4-\lambda}{\lambda}} \ , \
\lambda \in [0,4] \ .
\end{equation}
Repeating all steps as in the previous cases we find
\begin{equation}
\phi_{c^2}(z) = \frac{\sqrt{2}}{\sqrt{\sqrt{1-16z} + 1}} - 1 \ ,
\end{equation}
and the S-transform
\begin{equation}
S_{c^2}(z) = \frac{1}{4z} \left( \frac{1}{z+1} - \frac{1}{(z+1)^3} \right) \ .
\end{equation}
For this S-transform the cumulative distribution function for $C$ (\ref{eq:HLtheorem}) is given by the solution of the following equation:
\begin{equation}
4F^3_C(x)- x^2 F_C(x) - x^2 = 0 \ .
\end{equation}
This equation can be solved for $F_C(x)$. The solution has three branches, and one has to choose the branch that gives a monotonic function on $x \in [0,\sqrt{2}]$ increasing from
$F_C(x=0)=0$ to $F_C(x=\sqrt{2})=1$. The radial profile of the eigenvalue distribution is obtained by differentiating the solution $\varrho_C(r) = \frac{1}{2\pi r} F'_C(r)$. It is shown in Fig. (\ref{fig:SingleMatrices}c).

We additionally consider an isotropic random matrix $D=du$ constructed from an invariant Hermitian random matrix $d$ that has a uniform distribution $\rho_d(\lambda)=1$ for $\lambda\in [0,1]$. The radial profile of the distribution for the matrix $D$ is shown in Fig. (\ref{fig:SingleMatrices}d). 

In Section \ref{sec:fsp} we employ a finite size version of the aforementioned classes of random matrices to discuss products of large but finite isotropic matrices.

\section{Singular values of isotropic matrices}
\label{sec:sv}

As follows from Eq. (\ref{eq:HLtheorem}), the eigenvalues statistics of infinitely large isotropic matrices is in one-to-one correspondence with the statistics of singular values. Indeed, Eq. (\ref{eq:HLtheorem}) provides a relation between the eigenvalue density of an isotropic matrix $F_H(x)$ and the S-transform for the matrix $h^2=HH^\dagger$. Obviously, eigenvalues of $h$ are equal to singular values of $H$. From the S-transform one can derive the eigenvalue density of $h^2$.
For instance, for the Ginibre ensemble the eigenvalues are distributed uniformly on the unit disk. The eigenvalue density is $\rho(z) = 1/\pi$ for $|z|\le 1$ and hence  $F_H(x) = x^2$ for $x\in [0,1]$. Inserting this result into Eq. (\ref{eq:HLtheorem}) one finds $S_{h^2}(z) = 1/(1+z)$. This also means that the eigenvalue distribution of $h^2=HH^\dagger$ is given by the Wishart distribution (\ref{eq:W}) and the one of $h$ by the quarter-circle law (\ref{eq:QC}). This is of course the same calculation as the one presented in the previous section for matrix $A$, but now it has been carried out in the opposite direction.

An interesting situation is encountered for the matrix $P= HH\ldots H$ being a product of $n$ independent infinitely large Ginibre matrices $H$. As discussed in Section \ref{sec:prod}, $F_P(x) = F_H(x^{1/n}) = x^{2/n}$ (see Ref. \cite{bns}), and hence 
\begin{equation}
\label{eq:EVn}
\rho_P(z) = \frac{1}{2\pi|z|} F_P'(|z|) = \frac{1}{n\pi} |z|^{-\frac{2(n-1)}{n}}
\end{equation}
on the unit disk $|z|\le 1$. The S-transform for $p^2=PP^\dagger$ is
\begin{equation}
S_{p^2}(z) = \frac{1}{(1+z)^n} \ ,
\end{equation}
as follows from Eq. (\ref{eq:HLtheorem}). We can now write an explicit equation for the moment-generating function (\ref{eq:inverse})
\begin{equation}
\label{eq:FC}
z \left(1 + \phi_{p^2}(z)\right)^{n+1} = \phi_{p^2}(z) \ ,
\end{equation}
and for the Green's function (\ref{eq:phi}) 
\begin{equation}
\label{eq:Gn}
z^n G^{n+1}_{p^2}(z) = z G_{p^2}(z) - 1 \ .
\end{equation}
It is worth noting that one can derive an analogous equation also for the product of rectangular Gaussian random matrices \cite{bjlns1,bjlns2}. The solution of Eq. (\ref{eq:FC}) can be written as a power series 
\begin{equation}
\label{eq:moments}
\phi_{p^2}(z) =
\sum_{k=1}^\infty \mu_{k} z^k = \sum_{k=1}^\infty \frac{1}{(n+1)k+1} \binom{(n+1)k+1}{k} z^k \ ,
\end{equation} 
with the coefficients $\mu_k$ given by the Fuss-Catalan numbers \cite{gkp}. The radius of 
convergence of the power series $\phi_{p^2}(z)$ is equal to
\begin{equation}
\label{eq:rc}
r_c=\lim_{k\rightarrow \infty} \frac{\mu_{k}}{\mu_{k+1}} = 
\frac{n^n}{(n+1)^{n+1}} \ ,
\end{equation} 
and $\phi_{p^2}(z)$ is singular when $z$ approaches $r_c$.
The first moment of this distribution corresponds to the mean square of singular values of the matrix $P$: $\langle \lambda^2_{SV} \rangle_P = \mu_{1} = 1$.
This can be compared with the mean squared absolute value of eigenvalues of the matrix $P$, which is given by the second absolute moment of the distribution (\ref{eq:EVn}): $\langle |\lambda_{EV}|^2 \rangle_P = \int |z|^2 \rho_P(z) d^2z = 1/(n+1)$. We see that $\langle |\lambda_{EV}|^2 \rangle_P \le \langle \lambda^2_{SV} \rangle_P$, as expected
on general grounds.

Let us now discuss the behavior of the singular value density of $P$. Singular values of $P$ are equal to eigenvalues of the Hermitian matrix $p$, so this density is equal to the eigenvalue density $\rho_p(\lambda)$. First, let us determine the behavior of $\rho_{p^2}(\lambda)$ for $\lambda \rightarrow 0$. As follows from Eq. (\ref{eq:Gn}), the Green's function has a singularity for $z\rightarrow 0$:
\begin{equation}
G_{p^2}(z) \sim z^{-\frac{n}{n+1}} \ .
\end{equation}
From Eq. (\ref{eq:G}), this means that also the eigenvalue distribution must have the same singularity 
\begin{equation}
\rho_{p^2}(\lambda) \sim \lambda^{-\frac{n}{n+1}} 
\end{equation}
when $\lambda \rightarrow 0$, which corresponds to the following 
singularity of the eigenvalue density of $p$:
\begin{equation}
\label{eq:SVn}
\rho_p(\lambda) \sim \lambda^{-\frac{n-1}{n+1}} \ .
\end{equation}
The powers in Eq. (\ref{eq:EVn}) and Eq. (\ref{eq:SVn}) are related to each other. 
For $n=1$ both functions are regular at zero (as they are given by the quarter-circle and Ginibre laws, Eqs. (\ref{eq:QC}) and (\ref{eq:Ginibre}), respectively). For $n=2$ the eigenvalue density behaves as $|z|^{-1}$ while the singular value density as $\lambda^{-1/3}$, for $n=3$ they behave as $|z|^{-4/3}$ and $\lambda^{-1/2}$, respectively, etc. One can repeat the discussion also for products of rectangular matrices \cite{bjlns1,bjlns2}.

The function $G_{p^2}(z)$ has a cut along the interval $[0,(n+1)^{n+1}/n^n]$ on the real axis that corresponds to the support of the eigenvalue distribution $\rho_{h^2}(\lambda)$. The upper end of this interval is equal to $1/r_c$ (\ref{eq:rc}) as follows from the change of argument $z\rightarrow 1/z$ in $G(z)$ and $\phi(z)$ (see Eq. (\ref{eq:phi})). The position of the upper end of the interval can also be found directly from Eq. (\ref{eq:Gn}) as a place were the function $G=G_{p^2}(z)$ is singular. Singularity means that either $dG/dz=0$ or $dz/dG=0$. Writing Eq. (\ref{eq:Gn}) as 
$z^n G^{n+1} = z G - 1$ and differentiating both sides with respect to $G$ and setting $dz/dG=0$ we obtain $(n+1) z^n G^n = z$. Solving these two equations we find $z=(n+1)^{n+1}/n^n$, that corresponds to the singularity at the upper end of the support of the density $\rho_{p^2}(\lambda)$. One can also determine a 
closed-form expression for $\rho_{p^2}(\lambda)$ in terms of special functions \cite{pz}.

The relation of the singularities for $\lambda\rightarrow 0$ is actually more general. 
For any infinitely dimensional isotropic matrix $H$ whose eigenvalue density has a power-law singularity 
\begin{equation}
\rho_H(z) \sim |z|^{-s}
\end{equation} 
with $0<s<2$ we can determine the power of the corresponding singularity of the singular value density. From Eq. (\ref{eq:HLtheorem}) we can derive the behavior of the S-transform for $h^2$ for $z\rightarrow 0$ as $S_{h^2}(z) \sim (1+z)^{-2/(2-s)}$, which implies (see Eqs. (\ref{eq:phi})-(\ref{eq:inverse})) that the Green's function of $h^2$ has a singularity $G_{h^2}(z) \sim z^{-2/(4-s)}$ at the origin. This singularity is linked to the singularity of the eigenvalue density  $\rho_{h^2}(\lambda) \sim \lambda^{-2/(4-s)}$ for $\lambda\rightarrow 0$. By changing variables $\lambda \rightarrow \lambda^2$ we obtain a singularity of the eigenvalue density of $h$ (which is the density of singular values of $H$):
\begin{equation}
\rho_h(\lambda) \sim \lambda^{-\frac{s}{4-s}} \ .
\end{equation}
To summarize, an isotropic matrix whose eigenvalue density has a singularity $|z|^{-s}$ at the origin of the complex plane, has a singularity $\lambda^{-s/(4-s)}$ for $\lambda\rightarrow 0$ in the singular value density. This statement can be inverted. An infinitely large isotropic matrix $H$ having a power-law singularity in the density of singular values $\rho_h(\lambda) \sim \lambda^{-\sigma}$ with $0<\sigma<1$ has a power law singularity in the eigenvalue density with the following power
\begin{equation}
\rho_H(z) \sim |z|^{-\frac{4\sigma}{1+\sigma}} \ .
\end{equation} 

\section{Finite size effects}
\label{sec:fs}

So far we have discussed the limiting densities for $N\rightarrow \infty$. However, in many practical problems one encounters large but finite matrices. The calculation of the eigenvalue density for finite $N$ is much more complicated. Moreover, contrary to the limit $N\rightarrow \infty$, the results for finite $N$ are not universal and they depend on many details of the probability measure. The finite $N$-density has been calculated analytically only for a couple of very specific cases including Ginibre matrices \cite{ks,g1,g2}, elliptic matrices \cite{g3}, unitary truncated matrices \cite{sz} and products of independent Ginibre matrices \cite{ab}.  

Generally, various classes of isotropic random matrices may have the same limiting eigenvalue density for $N\rightarrow \infty$ but completely different properties for finite $N$. In this section we discuss the three most common classes of isotropic random matrices. The first class is isotropic random matrices defined by the partition function
 \cite{fz,fsz}
\begin{equation}
\label{eq:type1}
Z = \int DH \exp \left ( - N {\rm Tr} \; V\left(H H^\dagger\right) \right ) \ ,
\end{equation}
with a potential $V(x)$ which is an $N$-independent polynomial (or power series) in $x$. The symbol $DH$ denotes the flat measure for $N\times N$ complex matrices. 
 
The second type are random matrices constructed as
\begin{equation}
\label{eq:type2}
H' = h u
\end{equation}
where $u$ is an $N\times N$ Haar unitary matrix and $h$ is an $N\times N$ Hermitian matrix generated from an invariant ensemble defined by the partition function
\begin{equation}
\label{eq:type2_2}
Z = \int Dh \exp \left ( - N {\rm Tr} \; v\left(h\right) \right) \ .
\end{equation}
The potential $v(x)=v(-x)$ is an $N$-independent even polynomial (or power series) in $x$ and $Dh$ is the flat measure for $N\times N$ Hermitian matrices.  

Finally, the third type are random matrices constructed as
\begin{equation}
\label{eq:type3}
H''= u_1 h u_2 \ ,
\end{equation}
where $u_1$ and $u_2$ are independent matrices uniformly distributed (according to the Haar measure) on the unitary group $U(N)$, and $h$ is an $N\times N$ diagonal Hermitian matrix. The entries on the diagonal are independent identically distributed real (non-negative) random variables with an $N$-independent probability distribution, whose density we denote by $p(x)$. As far as the eigenvalue spectrum is concerned random matrices (\ref{eq:type3}) have the same eigenvalues as random matrices defined as a left multiplication of a diagonal Hermitian matrix by a Haar unitary matrix $hu$. Such matrices were studied in Ref. \cite{b} where they were called sub-unitary.

In all three cases the probability measures are invariant under the left $(H\rightarrow UH)$ and right $(H\rightarrow HU)$ multiplications by an arbitrary unitary matrix $U$.  A common feature of these ensembles is also that the functions $p,v,V$, which define the probability measures, do not depend on $N$. Otherwise, the three ensembles are completely different and have different eigenvalue statistics. For the matrices of the third type the eigenvalues are independent while for the other two they are not \cite{ks}. 

For the sake of illustration let us discuss three ensembles of isotropic matrices, all having the limiting density for $N\rightarrow \infty$: $\rho_H(z)=1/\pi$ on the unit disc. The matrix $H$ of the first type (\ref{eq:type1}) is defined by the partition function with the potential $V(x)=x$. It is a Ginibre matrix \cite{g1}. The matrix $H'$ of the second type is defined by the partition function with the potential $v(x)=x^2/2$. The matrix $H''$ of the third type (\ref{eq:type3}) is defined by the probability distribution with the probability density function given by the quarter-circle law $p(x) = \sqrt{4-x^2}/\pi$ for $x \in [0,2]$. In the limit $N\rightarrow \infty$ the eigenvalue densities of $H$, $H'$ and $H''$ tend to the same limiting distribution equal to $1/\pi$ on the unit disc, but for finite $N$ the eigenvalue densities of $H$, $H'$ and $H''$ are different. For $N=1$ they are 
\begin{equation}
\label{eq:rho}
\rho(z) = \frac{1}{\pi} e^{-|z|^2} \ ,
\end{equation}
for the first type,
\begin{equation}
\label{eq:rho_prime}
\rho'(z) = \frac{1}{\sqrt{2}\pi^{3/2} |z|} e^{-|z|^2/2} \ ,
\end{equation}
for the second one, and
\begin{equation}
\label{eq:rho_second}
\rho''(z) = \frac{\sqrt{4-|z|^2}}{2\pi^2 |z|}
\end{equation}
for the third one, respectively. The three expressions can be easily derived since for $N=1$ random matrices reduce to scalar random variables. The characteristic $1/|z|$ behavior for $H'$ and $H''$ is generated by the Jacobian of the transformation to polar coordinates of the flat measure $d^2 z = r dr d\phi $: since $\phi$ and $r$ are independent random variables we have $ p(r) dr \frac{1}{2\pi} d\phi =  
p(|z|) \frac{1}{2\pi|z|} d^2 z$. 

\begin{figure}
\centering
\includegraphics[width=0.3\textwidth]{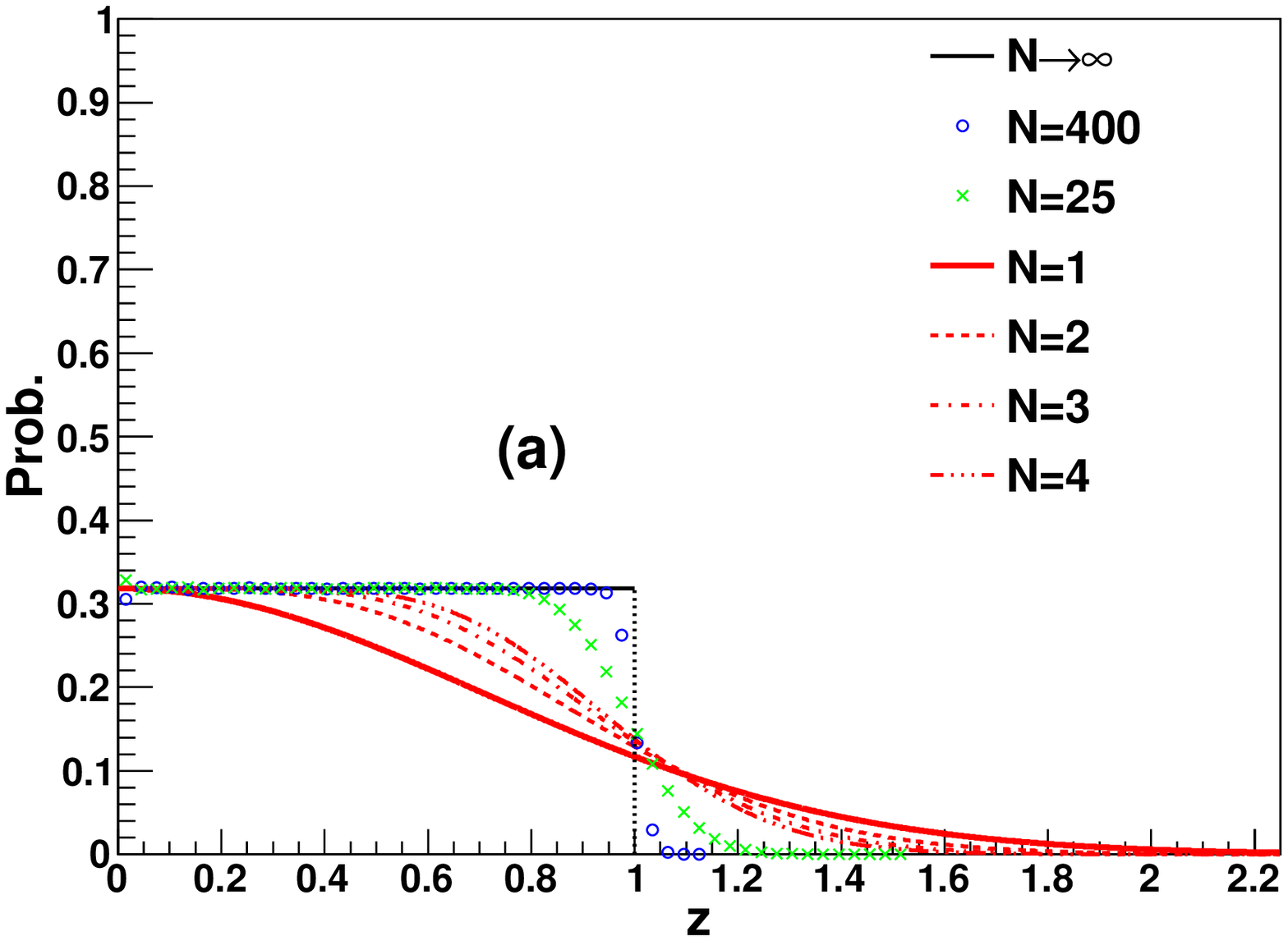}
\includegraphics[width=0.3\textwidth]{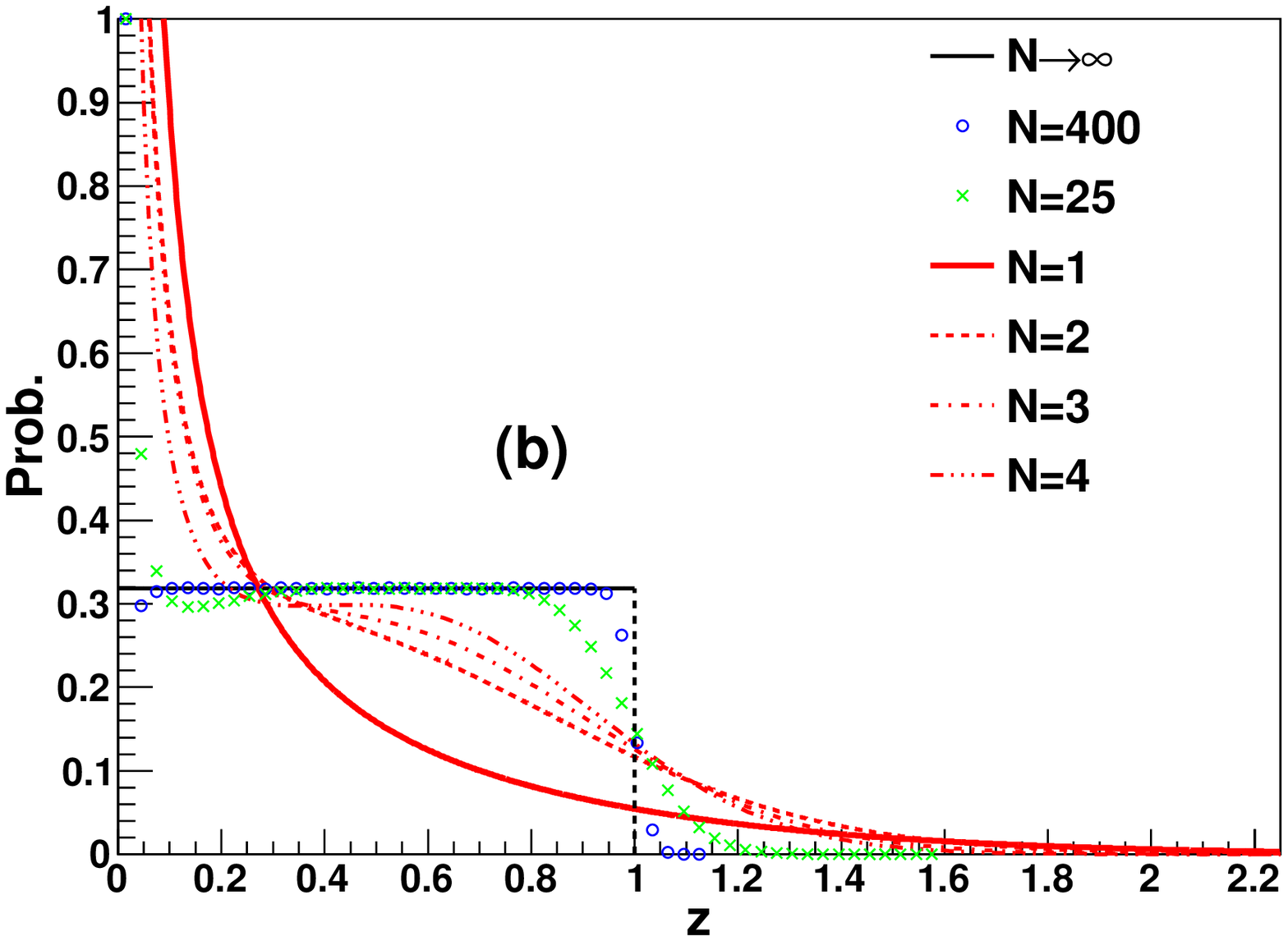}
\includegraphics[width=0.3\textwidth]{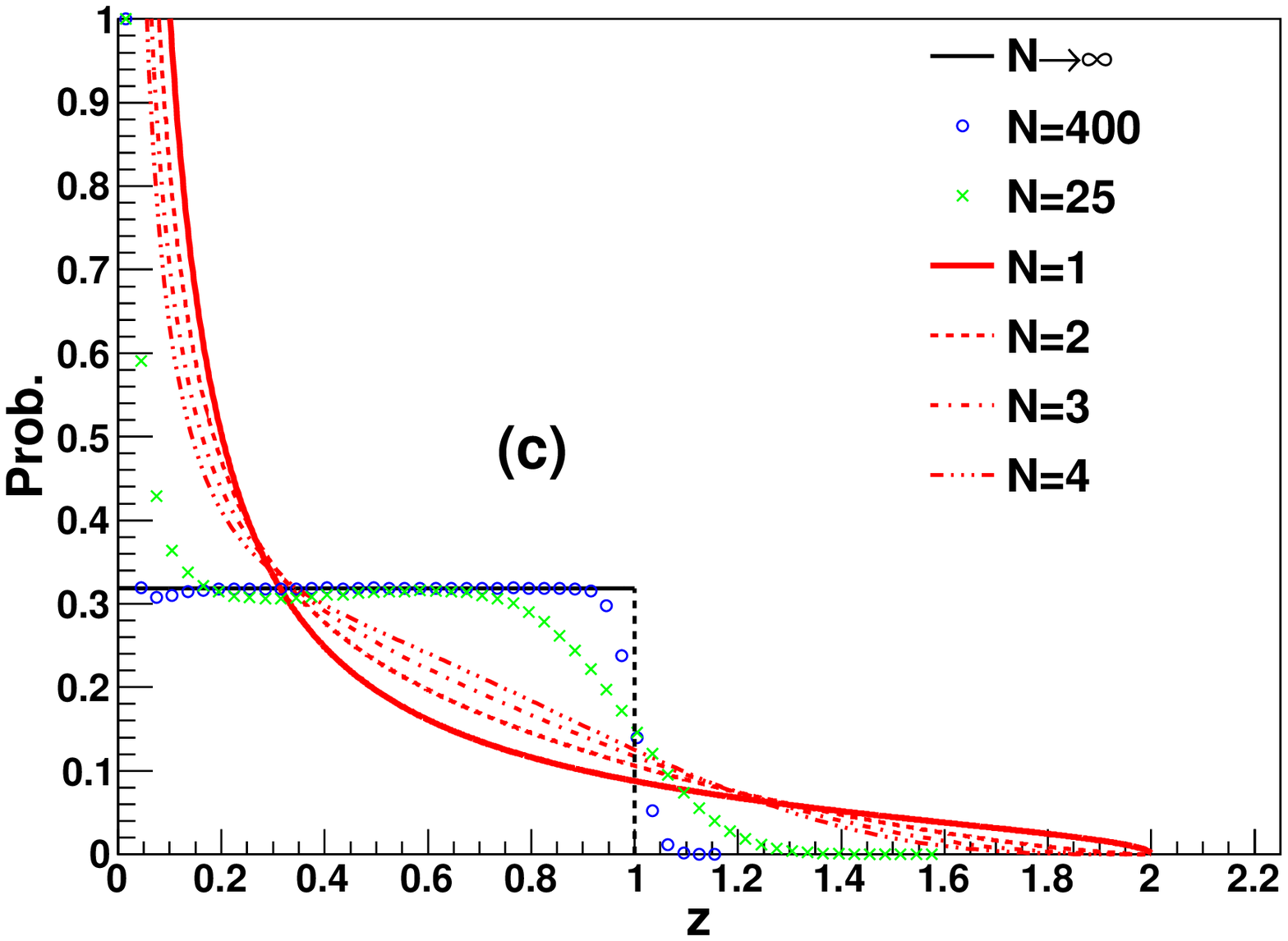}
\caption{(a) Numerical eigenvalue densities (for matrix size $N = 25$ (green crosses) and $400$ (blue circles), made of $10^{7}$ eigenvalues) for Ginibre matrices described by Eq. (\ref{eq:type1}), compared with the corresponding $N = 1$ (Eq. (\ref{eq:rho}), (red solid line), $N=2$ (Eq. (\ref{eq:rhoN_t1})) (red dashed line), $N=3$ (red dashed-dotted line), $N=4$ (red dashed-triple dotted line) and $N \rightarrow \infty$ (black line) densities. (b) Same plot as (a) for matrices of the type described by Eqs. (\ref{eq:type2}) and (\ref{eq:type2_2}). (c) Same plot as (a) for matrices of the type described by Eq. (\ref{eq:type3}).}
\label{fig:FiniteSize}
\end{figure}

One can also calculate the eigenvalue densities for $N=2,3,4,\ldots$. The details of the calculations are given in Appendix \ref{appendixA}. Except for the Gaussian ensemble of type one (\ref{eq:type1}), where a closed formula for finite $N$ density can be given in a simple form (\ref{eq:rhoN_t1}), the calculations for two remaining types of matrices get more and more tedious and cumbersome with increasing $N$. They become easy again for large $N$, for which one can anticipate a form of finite size corrections to the limiting density. In Fig. \ref{fig:FiniteSize} we show densities for $N=1,2,3,4,\infty$ which are obtained analytically and for $N=25,400$ which are obtained by Monte Carlo simulations. The finite-size corrections close to the edge of the support of the limiting density at $\lambda=1$ take a form of a sigmoidal function. The extent of the crossover region of this function shrinks with $N$ to zero in the limit $N\rightarrow \infty$. For the Gaussian ensemble of type (\ref{eq:type1}) the shape of the sigmoidal function is known to be given by a complementary error-function with the range parameter scaling as $1/\sqrt{N}$ \cite{ks}. Also for the two remaining types of finite-size ensembles (\ref{eq:type2}) and (\ref{eq:type3}) one expects the crossover behavior to be controlled by an error-function with the same $1/\sqrt{N}$-scaling \cite{b}. Indeed this is what we observe numerically. 

There is a significant difference between the approach of the finite $N$ densities to the limiting density at zero for the first type of matrices (\ref{eq:type1}) for which the approach is uniform and the two remaining types (\ref{eq:type2},\ref{eq:type3}) for which it is non-uniform (see Fig. \ref{fig:FiniteSize}). The non-uniform behavior is a remnant of the $1/|z|$ singularity in Eqs. (\ref{eq:rho_prime}) and (\ref{eq:rho_second}). When $N$ increases the singularity is pushed towards zero and it eventually disappears in the limit. To compensate for the excess of eigenvalues in the region close to the origin the finite $N$ profiles develop a shallow dip for intermediate values of $\lambda$ as can be seen by eye for $N=25$ and $N=400$ in Figs. \ref{fig:FiniteSize}.b and \ref{fig:FiniteSize}.c. The effect of the excess of eigenvalues at the origin of the complex plane is not present for matrices of type one (\ref{eq:type1}) because of the repulsion of eigenvalues from the origin \cite{ghs}.  

\section{Products of finite matrices}
\label{sec:fsp}

\begin{figure}
\centering
\includegraphics[width=0.3\textwidth]{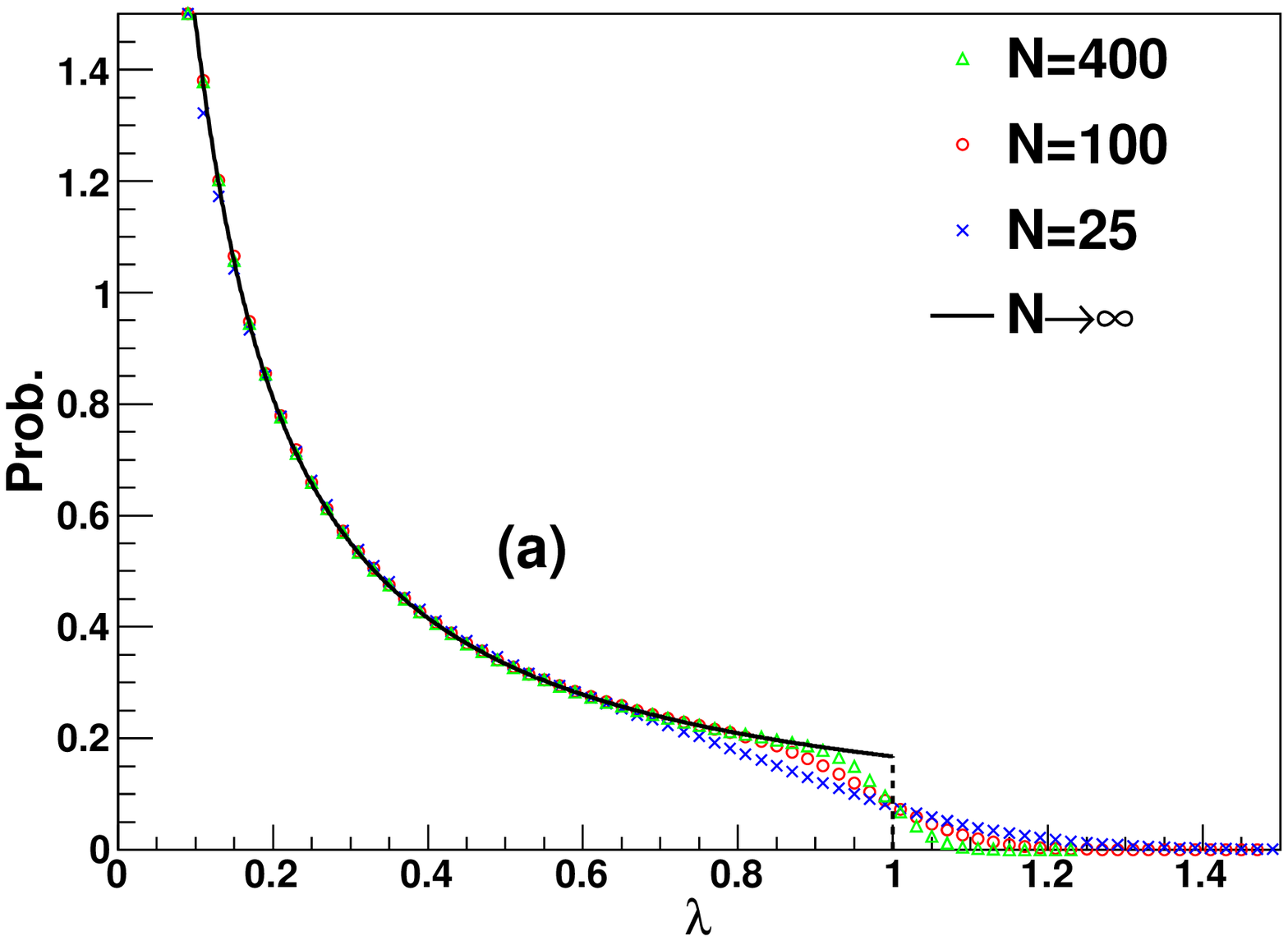}
\includegraphics[width=0.3\textwidth]{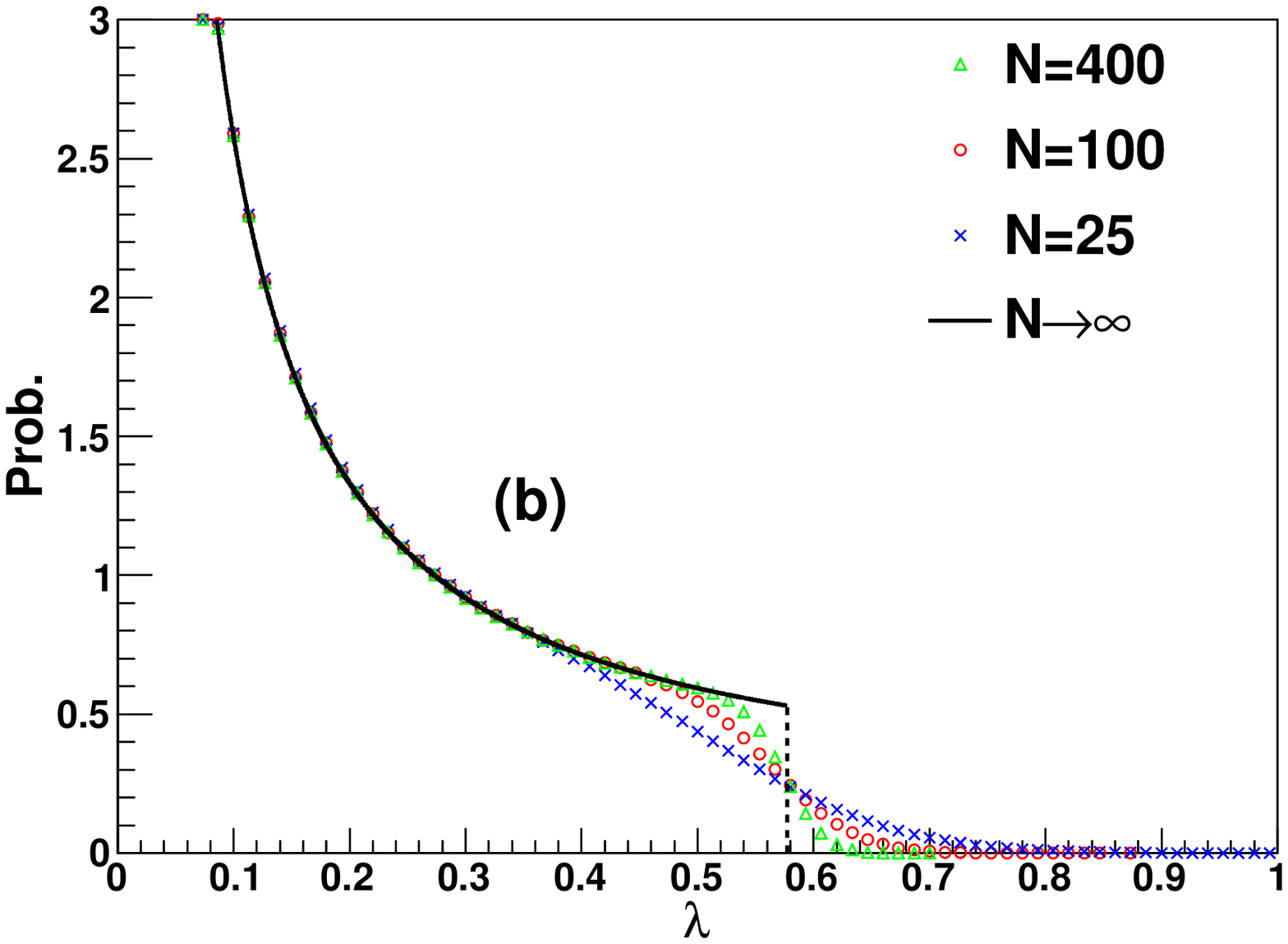}
\includegraphics[width=0.3\textwidth]{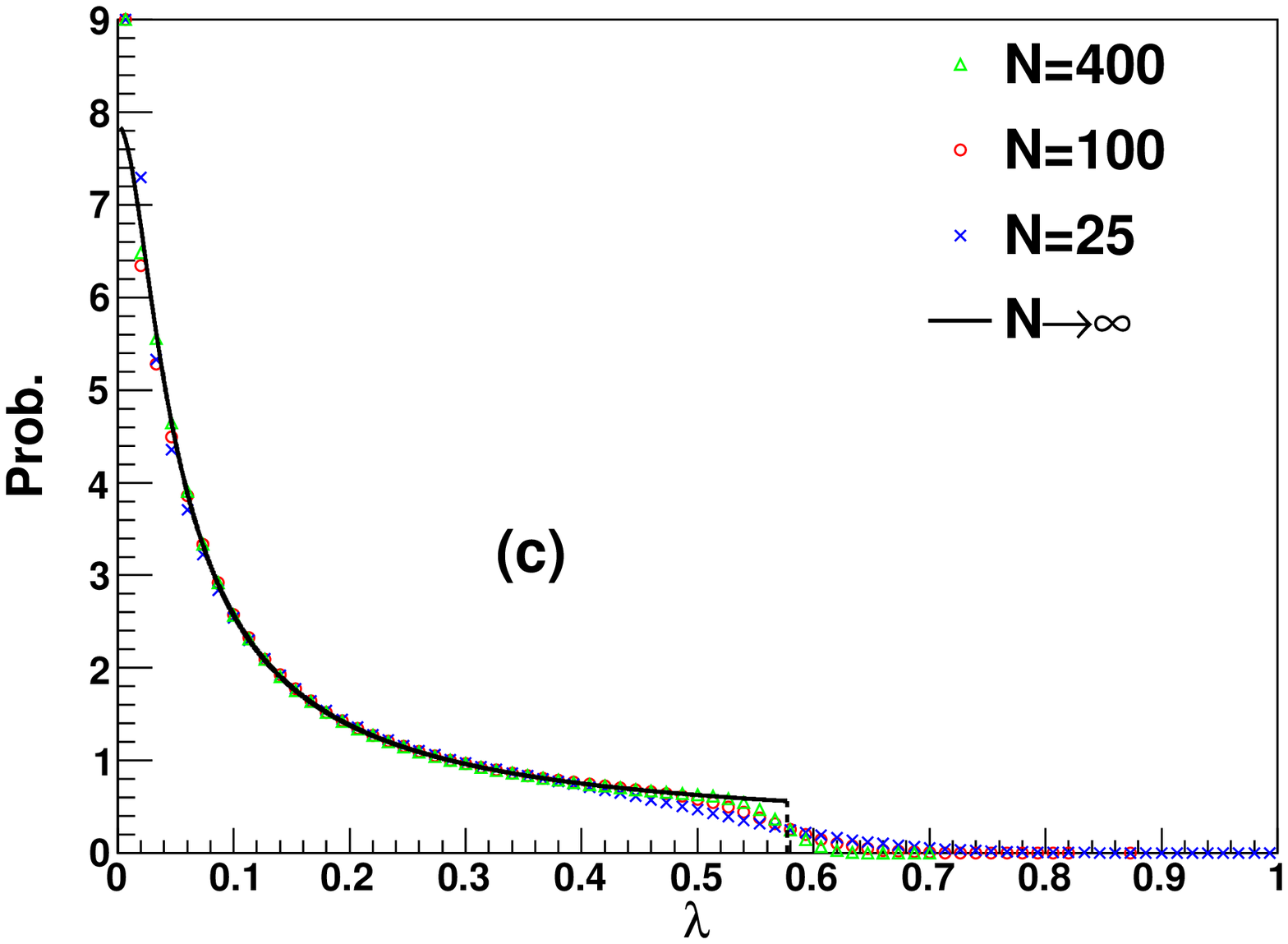}
\caption{Monte Carlo eigenvalue densities of products of two isotropic random matrices compared with the theoretical predictions (black lines) obtained from the solution of Eq. (\ref{eq:S1n}). Different plots refer to the products $AB$ (a), $AD$ (b), and $BD$ (c). In all three cases, the results obtained with three different matrix sizes ($N = 25$ (blue crosses), $100$ (red circles) and $400$ (green triangles)) are shown. All Monte Carlo densities are made of $10^7$ eigenvalues.}
\label{fig:SimpleProducts}
\end{figure}

In this section we study how the commutative and self-averaging properties of finite isotropic matrices set in when $N$ increases. To this end we exploit finite size versions of $A,B,C,D$ matrices introduced in Section \ref{sec:examples}. As $A$ we take Ginibre matrices \cite{g1} which belong to the first class (\ref{eq:type1}) discussed in the previous section, while as $B,C,D$ we take matrices (\ref{eq:type3}) constructed from diagonal random matrices by isotropic unitary randomization. First we study the size dependence of the eigenvalue densities for products of two matrices 
similarly as we did for single matrices in Section \ref{sec:fs}. The finite $N$ spectra are generated by Monte-Carlo simulations and they are compared in Fig. \ref{fig:SimpleProducts} to the corresponding limiting densities for $N\rightarrow \infty$ which were obtained from the $S$-transform manipulations and the Haagerup-Larsen theorem, as described in Section \ref{sec:prod}). The size dependence of the finite $N$ densities is very weak in the bulk of the distribution as one can see in Fig.\ref{fig:SimpleProducts}, where the data points for $N=25$, $N=100$ and $N=400$ lie on top of the limiting curve. A significant dependence on $N$ is observed only in the region close to the edge of the distribution. In this region the density takes the form of a sigmoidal function. The range of this function shrinks when $N$ increases
to eventually restore a sharp threshold at the edge in the limit $N\rightarrow \infty$. This effect is analogous to that discussed for products of Gaussian matrices in Refs. \cite{bjlns1,bjlns2,bjw}, where it was conjectured that the sigmoidal function in that case was given by the complementary error function. 

\begin{figure}
\centering
\includegraphics[width=0.3\textwidth]{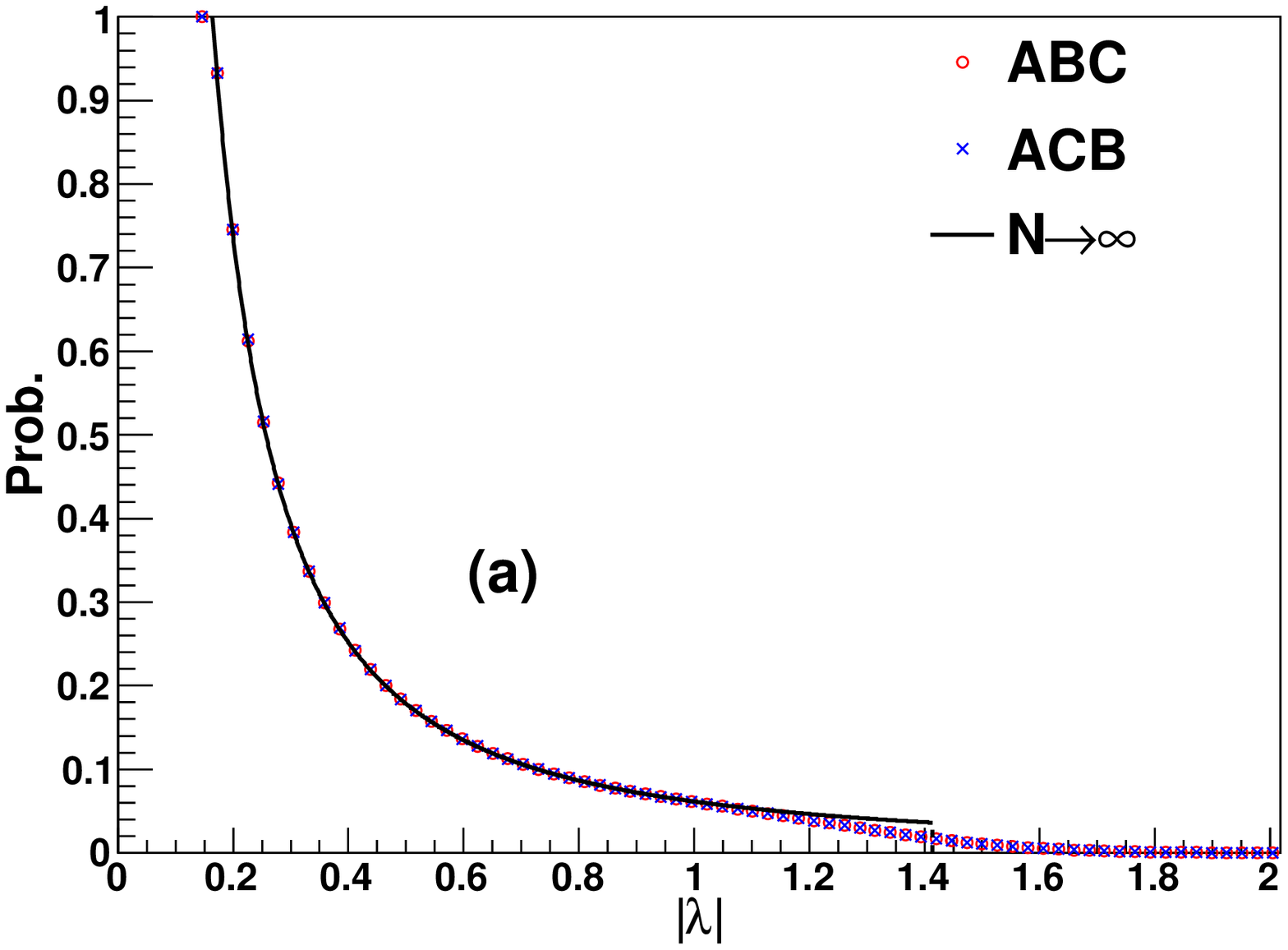}
\includegraphics[width=0.3\textwidth]{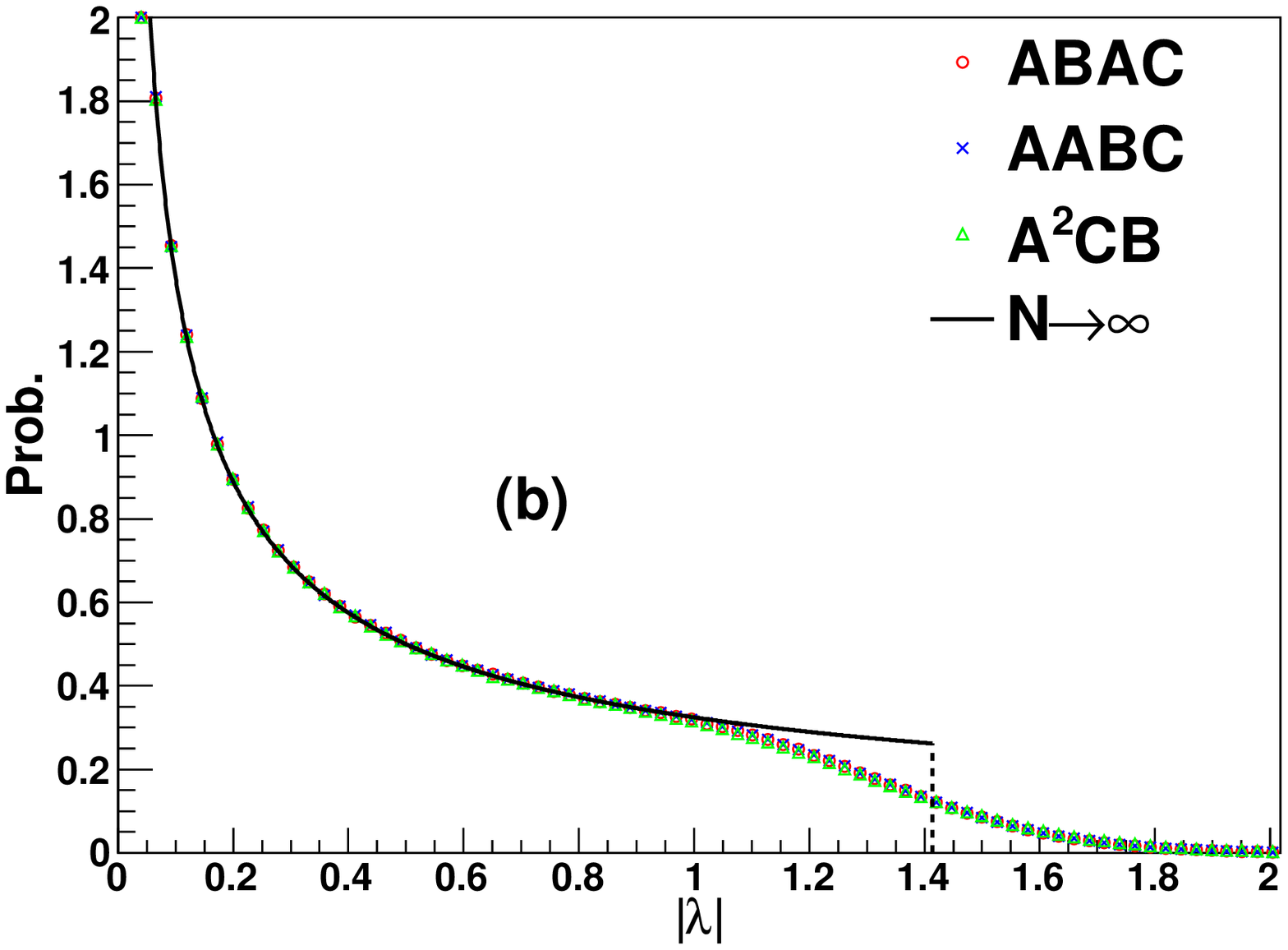}
\includegraphics[width=0.3\textwidth]{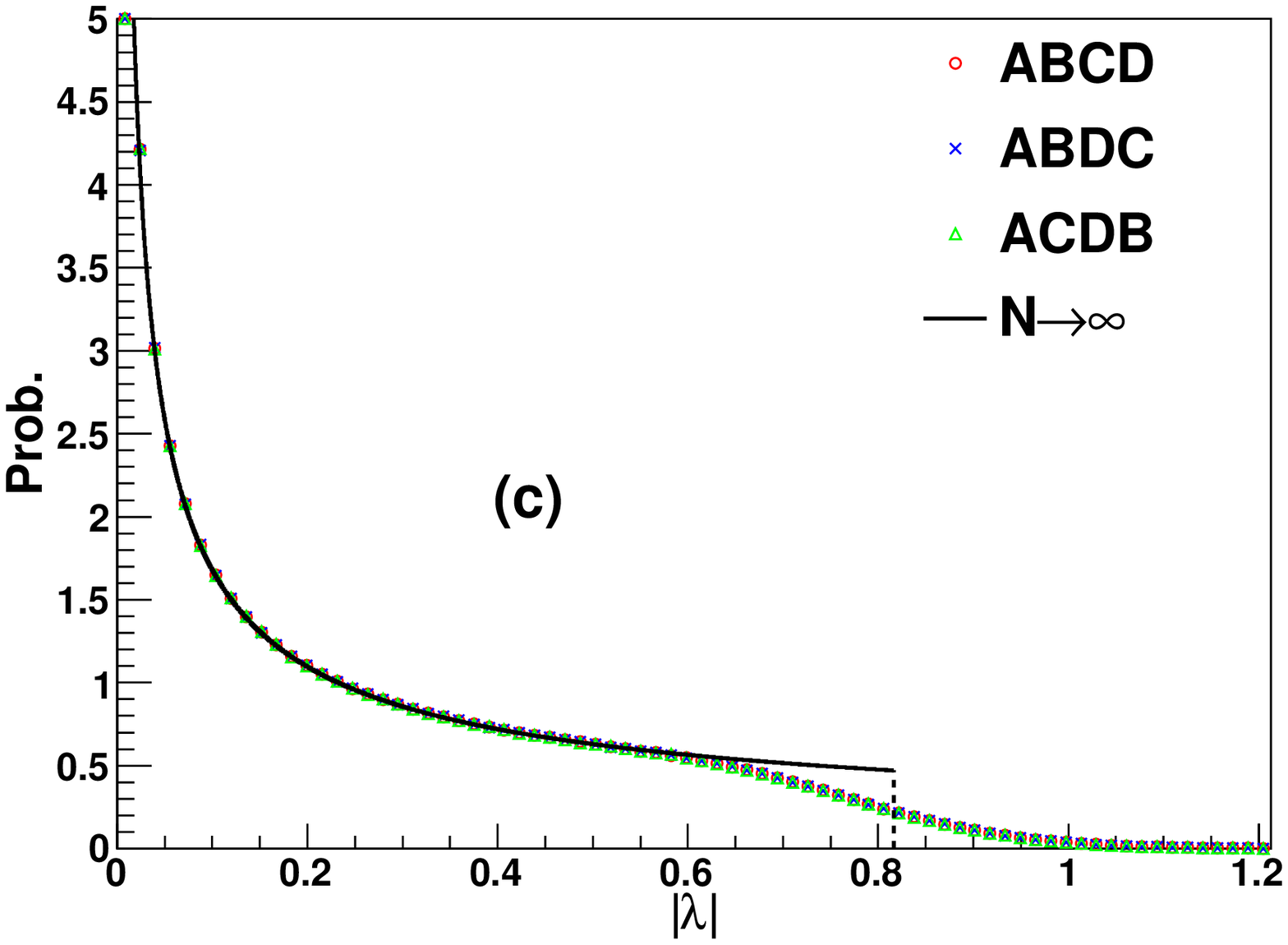}
\caption{Monte Carlo eigenvalue densities of products of isotropic random matrices. Different plots refer to the products: (a) $ABC$ (red circles) and $ACB$ (blue crosses), (b) $ABAC$ (red circles), $AABC$ (blue circles) and $A^2CB$ (green triangles), and (c) $ABCD$ (red circles), $ABDC$ (blue crosses) and $ACDB$ (green triangles). In all plots black lines represent the limiting result ($N \rightarrow \infty$) obtained as solution of Eq. (\ref{eq:S1n}). All Monte Carlo densities have been produced by numerical diagonalization of $10^{5}$ matrices of size $N=100$.}
\label{fig:OrderInvariance}
\end{figure}

In turn, in Fig. \ref{fig:OrderInvariance} we restrict ourselves to matrices with $N=100$ but we compare densities for products of isotropic random matrices multiplied in  different order, as for instance $ABDC$ and $ACDB$. We see on each plot in Fig. \ref{fig:OrderInvariance} that data points representing different order of multiplication lie on top of a master curve within the symbol size. This means that already for matrices of size of order $N=100$ the multiplication of isotropic random matrices can be treated in practical applications as spectrally commutative.
 
We also check self-averaging by comparing products containing multiple uses of a single matrix to products of independent matrices. For example, we consider a product of the type $A^2 BC$ and $AABC$, in the first of which $A$ is used twice and in the second of which two different $A$'s, representing identically distributed but independent matrices, are used. Again the deviations between the resulting spectra are small and practically not detectable in the observed resolution (see Fig. \ref{fig:OrderInvariance}.b). However, one can generally observe in the finite $N$ eigenvalue spectra that the multiple use of a single matrix in a product has a stronger influence on the shape of the spectrum than the change of matrix ordering in this product. The effect has been studied quantitatively only for products of Ginibre matrices, for which one can analytically derive a finite $N$ eigenvalue density for the product of $n$ independent matrices $AA\ldots A$ and for the corresponding power $A^n$ of a single matrix \cite{ab}. In this case one explicitly sees that the range of the sigmoidal function corrections at the edge of the spectrum is of order $a/\sqrt{N}$ with a coefficient $a$ that increases as $\sqrt{n}$ for the product of independent matrices and as $n$ for the $n$-th power. Thus the finite size effects are bigger in the latter case.

\section{Discussion}
\label{sec:discussion}

We have shown that the eigenvalue densities of products of isotropic random matrices do not depend, in the large $N$ limit, on the multiplication ordering. They only depend on the random matrix ensembles from which the matrices in the product are generated. We have also extended the result on self-averaging \cite{bns} by showing that a single matrix from an isotropic random matrix ensemble is representative enough to describe multiple independent occurrences of matrices from the same ensemble within the same matrix product. 
We have also derived a relation between the exponents which determine the singular behavior at zero of the eigenvalue density and the singular value density of infinitely large isotropic random matrices. This result generalizes a previously known relation for products of Gaussian matrices \cite{bjlns1,bjlns2,bjw}.

Isotropic random matrices are a very special class of non-Hermitian random matrix ensembles. Commutative and self-averaging properties of products of such matrices in the large $N$ limit follow from the Haagerup-Larsen theorem \cite{hl} and the existence of a one-to-one correspondence between invariant large matrices and free random variables, as well as from the commutative and associative properties of the multiplication law (\ref{eq:Sab}). It would be very interesting to work out similar relations for products of generic non-Hermitian matrices. The first step in this direction has been done in Ref. \cite{bjn} where the corresponding multiplication law for a large class of non-Hermitian matrices has been derived in the large $N$ limit. The generalization of the R and S transforms leads in that case to quaternion-valued functions of quaternion-valued arguments. The corresponding multiplication law is not commutative and thus the problem is more complicated. 

\appendix

\section{Calculations of eigenvalue density for finite $N$}
\label{appendixA}

In this appendix we discuss how to calculate eigenvalue distributions for finite $N$ isotropic random matrices of the three types of ensembles introduced in section (\ref{sec:fs}). For the sake of illustration we concentrate on matrices that for $N\rightarrow \infty$ have the limiting density $\rho(z)=1/\pi$ on the unit. 

For matrices of the type (\ref{eq:type1}) the corresponding matrix $H$ is an $N \times N$ complex matrix. The partition function has in this case a linear potential $V(x)=x$
\begin{equation}
Z = \int DH e^{- N {\rm Tr} \; H H^\dagger} \ .
\end{equation}
This is the standard Ginibre ensemble \cite{g1}. Matrices of this type have been thoroughly studied in the literature. Here we recall the main results and refer the reader to Ref. (\cite{ks}) for details. As it stands, the integrand defining the partition function depends on $N^2$ complex degrees of freedom corresponding to the matrix elements. When one is interested only in quantities depending on the eigenvalues, one can reduce the complexity of the problem by leaving an explicit dependence only on the $N$ complex eigenvalues $z_i$, $i=1,\ldots, N$ of the matrix by integrating out the remaining degrees of freedom. This gives, up to a normalization constant:
\begin{equation}
Z = \int dz^2_1 \ldots dz^2_N \; P(z_1,\ldots,z_N)  \ ,
\end{equation}
where $P(z_1,\ldots,z_N)$ is the eigenvalue joint probability density function 
\begin{equation}
P(z_1,\ldots,z_N) = \frac{N^{N(N+1)/2}}{\pi^{N}\prod_{k=1}^N k!} \; e^{-N\sum_{n=1}^N |z_n|^2} 
\prod_{1\le i < j \le N} |z_j-z_i|^2 \ .
\end{equation}
It is normalized in such a way that $\int dz^2_1 \ldots dz^2_N P(z_1,\ldots,z_N) =1$. Integrating all but one eigenvalue from the joint probability density function one obtains the eigenvalue density of $H$ for finite $N$
\begin{equation}
\rho_N(z) = \int dz_2^2 \ldots dz_N^2 \; P(z,z_2,\ldots,z_N) \ .
\end{equation}
Note that $P(z_1,\ldots,z_N)$ is symmetric with respect to permutations of eigenvalues.
Therefore it does not matter which $N-1$ eigenvalues are integrated out to obtain the density. The result reads 
\begin{equation}
\label{eq:rhoN_t1}
\rho_N(z) = e^{-N|z|^2} \sum_{n=0}^{N-1} \frac{(N|z|^2)^n}{n!} \ .
\end{equation} 
The radial profiles of these functions for different values of $N$ are plotted in Fig.\ref{fig:FiniteSize}.a. For large $N$ the shape of the radial profile is well approximated by a complementary error function changing from $0$ to $1$ in a region whose size scales like $1/\sqrt{N}$. In the limit $N\rightarrow \infty$ this region shrinks to one point, so the function becomes a step function. For a detailed discussion we again refer the reader to the excellent review \cite{ks}. 

Let us now discuss matrices of the third type (\ref{eq:type3}) constructed from diagonal matrices $h$ by unitary randomization $H=u_1 h u_2$ with two independent Haar unitary matrices $u_1$ and $u_2$. The ensemble of such matrices has the same eigenvalue density as the ensemble of matrices $H=hu$ randomized only on one side. The eigenvalue density for the latter one can be calculated as
\begin{equation}
\rho_H(z) = \left\langle \frac{1}{N}\sum_{i=1}^N \delta^{(2)}\left(z-\lambda_i\right)
\right\rangle_{h,u}
\end{equation}
where $\lambda_i$'s are eigenvalues of $H$, and the average is taken over $h$ and $u$. One can first average over $u$'s and then over $h$'s:
\begin{equation}
\rho_H(z) = \int dh_1\ldots dh_N P(h_1,\ldots,p_N) \rho_{h_1,\ldots,h_N}(z) \ .
\label{eq:rhoHrho0}
\end{equation}
Here $\rho_{h_1,\ldots,h_N}(z)$ is the result of averaging over $u$ 
for a fixed $h=\mbox{diag}(h_1,\ldots,h_N)$:
\begin{equation}
\rho_{h_1,\ldots,h_N}(z) = \left\langle \frac{1}{N}\sum_{i=1}^N \delta^{(2)}\left(z-\lambda_i\right) \right\rangle_{u} \ .
\end{equation}
The integration measure $\int dh_1\ldots dh_N P(h_1,\ldots,h_N)$ is the probability measure for the diagonal matrix $h$. For independent and identically distributed $h_i$'s it factorizes as $P(h_1,\ldots.h_N) =p(h_1)\ldots p(h_N)$, where $p(x)$ is the probability density function for diagonal elements. In our case it is $p(x) = \sqrt{4-x^2}/\pi$ for $x\in [0,2]$, so we have
\begin{equation}
\rho_H(z) = \int_0^2 dh_1 \frac{\sqrt{4-h_1^2}}{\pi}
\ldots \int_0^2 dh_N \frac{\sqrt{4-h_N^2}}{\pi} \ \rho_{h_1,\ldots,h_N}(z) \ . 
\label{eq:rhoHrho1}
\end{equation}
The eigenvalue distribution $\rho_{h_1,\ldots,h_N}(z)$ has been explicitly calculated in Ref. \cite{wf}. In order to write down the result it is convenient to order the $h_i$'s: $0\le h_1< h_2 \ldots < h_N$. With this ordering the density $\rho_{h_1,\ldots,h_N}(z)$ is given by 
\begin{equation}
\label{eq:Psi}
\rho_{h_1\ldots h_N} (z) = \frac{1}{\pi N} \sum_{i = k+1}^N F(h_i,|z|) \ , \quad
h_k < |z| < h_{k+1} \
\end{equation}
with the $F$'s defined on rings $|z| \in (h_k,h_{k+1})$ for $k=1,\ldots, N-1$. 
Inside the disk or radius $h_1$ ($|z|\le h_1$) and outside the disk of radius $h_N$ ($|z|>h_N$) the density vanishes $\rho_{h_1\ldots h_N} (z)=0$.
The function $F$ reads
\begin{equation}
\label{eq:F}
F(h_i,|z|) = \frac{(h_i^2 - |z|^2)^{N-2}}{\prod_{j \neq i} (h_i^2 - h_j^2)} \sum_{\ell = 0}^{N-1} s_{[i]}^\ell |z|^{-2(\ell+1)} \frac{1}{\binom{N-1}{\ell}} \left [ \ell h_i^2 + (N-1-\ell) |z|^2 \right ] \ ,
\end{equation}
and the symbols $s_{[i]}^\ell$ are defined as symmetric polynomials of order $\ell$ in $h_{i}^{2}$ variables, putting $h_{i}=0$. In other words, one can write $s^0 = 1$, $s^1 = \sum_{i=1}^N h_i^2$, $s^2 = \sum_{i<j}^N h_i^2 h_j^2$ etc., and take $s^\ell_{[i]} = s^\ell |_{h_i = 0}$. For different orderings of the $h_i$'s, result can be mapped onto the one given above by an appropriate permutation of the indices that organizes the $h$'s in increasing order. Therefore it is sufficient to calculate the integral (\ref{eq:rhoHrho1}) only for $h_1< h_2 \ldots < h_N$ since it takes the same value for all remaining orderings (permutations):
\begin{equation}
\rho_H(z) = N! \int_0^2 \frac{dh_1}{\pi} \sqrt{4-h_1^2}
\ldots \int_{h_{N-1}}^2 \frac{dh_N}{\pi} \sqrt{4-h_N^2} \rho_{h_1,\ldots,h_N}(z) \ . 
\label{eq:rhoHrho2}
\end{equation} 
One can also give an explicit form of the function $F$ for the case when two or more $h_i$'s have the same values \cite{wf}, but this case is irrelevant from the point of view of the last integral, since it gives a contribution of measure zero. 

For illustration let us give an explicit form of $\rho_{h1,h2}(z)$ for $N=2$ that follows from (\ref{eq:F}). Assuming $h_1<h_2$ the density reads
\begin{equation}
\label{eq:wf2}
\rho(z) = \left \{ \begin{array}{ll}
\frac{1}{2\pi} \frac{1}{h_2^2 - h_1^2} \left ( 1 + \frac{h_1^2 h_2^2}{|z|^4} \right ) & h_1< |z| < h_2 \\
0 & \ \mbox{otherwise .} 
\end{array} \right .
\end{equation}
One can write explicit expressions also for $N=3,4,\ldots$ using (\ref{eq:F}) and integrate them over $h$'s (\ref{eq:rhoHrho2}). We have done that for $N=2,3,4$. The results are shown in Fig. \ref{fig:FiniteSize}.c.

Let us make a general remark. We see that already for finite $N$ the matrix $H=hu$, where $u$ is a Haar unitary matrix on $U(N)$ and $h$ is a constant matrix $h={\rm diag}(h_1,\ldots,h_N)$, has many properties that are expected in the large $N$ limit. It has a spherically symmetric eigenvalue density on a ring whose radii depend on $h$'s. Actually one can show \cite{b} that Eqs. (\ref{eq:Psi},\ref{eq:F}) reproduce the Haaregup-Larsen equation in the large $N$ limit, when the distribution of $h_i$'s becomes a continuous function. 

We can apply a similar strategy to the finite $N$ ensembles of matrices $H=hu$ of the second type (\ref{eq:type2}) where now $h$ is a Hermitian matrix from an invariant unitary
ensemble. We can use again Eq. (\ref{eq:rhoHrho0}) but with $P(h_1,\ldots,h_N)$ 
being the joint probability function  \cite{m}
\begin{equation}
P(h_1,\ldots, h_N) = C_N e^{-\frac{N}{2} \sum_{i=1}^N h_i^2} \prod_{j<k} (h_k-h_j)^2 \ ,
\end{equation}
with $C_N$ being a normalization constant such that $\int dh_1\ldots h_N P(h_1,\ldots, h_N)=1$. There are two essential differences with respect to the previous case: the joint probability $P$ cannot be factorized, and the arguments $h_i$ of $P$ may take negative values. While doing the integration in 
(\ref{eq:rhoHrho0}) over $h$'s, it is convenient to restrict to non-negative semi-axes $h_i\ge 0$, for all $i=1,\ldots. N$. To this end we introduce a new function
$Q$ defined for non-negative $h$'s which is obtained from $P$
by integrating out signs of $h$'s:
\begin{equation}
Q(h_1,\ldots, h_N) = \sum_{s_1=\pm 1, \ldots s_N=\pm 1} 
P(s_1 h_1,\ldots, s_N h_N) \ .
\end{equation}
Since the density $\rho_{h_1\ldots h_N}(z)$ depends only on the absolute values of $h$'s ($\rho_{h_1\ldots h_N}(z) = \rho_{|h_1|\ldots |h_N|}(z)$) we have
\begin{equation}
\rho_H(z) = \int_{-\infty}^\infty dh_1\ldots   \int_{-\infty}^\infty  
dh_N P(h_1,\ldots,p_N) \rho_{h_1,\ldots,h_N}(z) = 
\int_{0}^\infty dh_1\ldots   \int_{0}^\infty  
dh_N Q(h_1,\ldots,h_N) \rho_{h_1,\ldots,h_N}(z) \ .
\end{equation}
For illustration, let us write it for $N=2$
\begin{equation}
Q(h_1,h_2) = 2 C_2 e^{-h_1^2-h_2^2} 
\left((h_2-h_1)^2 + (h_2+h_1)^2\right) \ . 
\end{equation}
The integral over all positive $h$'s can be now reduced to an integral over ordered sets $h_1<h_2<\ldots<h_N$ as before,
\begin{equation}
\rho_H(z) = N! \int_{0}^\infty dh_1\ldots   \int_{h_{N-1}}^\infty  
dh_N Q(h_1,\ldots,p_N) \rho_{h_1,\ldots,h_N}(z) \ ,
\end{equation} 
with $\rho_{h_1,\ldots,h_N}(z)$ given by Eq. (\ref{eq:Psi}). 
Using this method we have calculated $\rho_H(z)$ for $N=2,3,4$.
The result is presented in Fig. \ref{fig:FiniteSize}.b.

\subsection*{Acknowledgements}
We thank M.A. Nowak and R.A. Janik for interesting discussions and P. Vivo for interesting discussions and drawing our attention to the paper \cite{wf}.
G.L. acknowledges the Marian Smoluchowski Institute of Physics in Krakow for warm hospitality. Z.B. acknowledges financial support by the Grant DEC-2011/02/A/ST1/00119 of the National Centre of Science.

\end{document}